\documentclass[iop,apj,tighten]{emulateapj}
\usepackage{apjfonts,graphicx} 
\usepackage{amsmath,amstext}
\usepackage[breaklinks,colorlinks,citecolor=blue,linkcolor=blue]{hyperref} 
\usepackage{tabularx, graphicx}
\usepackage{xcolor}
\usepackage[all]{hypcap} 

\newcommand{\cname}{MACSJ0416.1-2403}

\newcommand{\hst}{{\it HST}}
\newcommand{\hff}{{\it HFF}}

\newcommand{\dls}{D_\mathrm{ls}}
\newcommand{\ds}{D_\mathrm{s}}

\newcommand{\vect}[1]{\boldsymbol{#1}}

\begin{document}

\title{An Evaluation of 10 Lensing Models of the Frontier Fields Cluster \cname}

\author{J.D. Remolina Gonz\'{a}lez\altaffilmark{1}, K. Sharon\altaffilmark{1}, \& G. Mahler\altaffilmark{1}}

\slugcomment{Accepted June 25, 2018 to ApJ}
\shorttitle{\hff\ V3 Lens Models of \cname}
\shortauthors{J. D. Remolina Gonz\'{a}lez, K. Sharon, \& G. Mahler}

\altaffiltext{1}{Department of Astronomy, University of Michigan, Ann Arbor, MI 48109, USA}
\altaffiltext{*}{jremolin@umich.edu}

\date{\today}

\begin{abstract}

Galaxy clusters can act as gravitational lenses to magnify the universe behind them, allowing us to see deep into the early universe. The {\it{Hubble Space Telescope}} {\it {Frontier Fields}} program \citep{Lotz:17} uses six galaxy clusters imaged by {\it{Hubble}} to discover and study galaxies at  $z\sim 5-10$. Seven independent teams developed lens models and derived magnifications for each galaxy cluster, based on positional and redshift constraints from the best available data at the time. In this work we evaluate ten models for \cname, which were made public in 2015, by contrasting them with new spectroscopic redshifts that were measured in 2016 \citep{Caminha:17}. We developed an independent comparison method that uses the source plane root-mean-square as a metric of lensing model performance. Our analysis quantifies the ability of models to predict unknown multiple images. We examine the source plane scatter of multiply-imaged systems and explore dependence of the scatter on the location and the redshift of the background sources. The analysis we present evaluates the performance of the different algorithms in the specific case of the \cname\ models.
\end{abstract}



\section{Introduction}
Strong gravitational lensing in massive galaxy clusters provides an opportunity to study both the physical properties of clusters, and the highly-magnified background universe behind them. 
Strong lensing analysis can directly inform estimates of the total (baryonic and dark matter) projected mass distribution of galaxy cluster cores (e.g., \citealt{Richard:10a,Sharon:14,Sharon:15b,Zitrin:16,Monna:17,Mahler:18}); the high magnifications associated with strong lensing clusters facilitate both observations the high redshift ($z\sim 6-11$) magnified background universe (e.g., \citealt{Kneib:04,Richard:08,Coe:13,Coe:15,Bouwens:14,Bradley:14,Atek:15,Salmon:17}), and  zoomed-in studies of highly-magnified galaxies around $z\sim 2$ (e.g., \citealt{Jones:10, Bayliss:11, Livermore:15, Johnson:17a, Johnson:17b}). Strong lensing clusters have also been used to constrain cosmological parameters (e.g., \citealt{Jullo:10,Cao:15,Magaña:15,Acebron:17}).
The {\it{Hubble Space Telescope}} (\hst) \ has been transformative for strong lensing research, enabling most of the studies listed above. 

The {\it{Hubble}} Frontier Fields (\hff; \citealt{Lotz:17}) targets six massive galaxy clusters used as cosmic telescopes to observe the background universe ($z \sim 5-10$). Each of the galaxy clusters and parallel fields was observed with seven optical and near-infrared bands using the Wide Field Camera 3 (WFC3) and the Advanced Camera for Surveys (ACS) over 140 orbits each for a total of 840 \hst\ orbits. Follow-up observations from a variety of observatories both on the ground and in space have provided complementary data, including spectroscopic redshifts ($z_{spec}$), which allowed for precise lens models and the magnification measurements needed to study the background universe (e.g., \citealt{Oesch:15,Alavi:16,Laporte:16,McLeod:16,Bouwens:17,Livermore:17,Yue:17,Atek:18,Kawamata:18}). \footnote{See \href{http://www.stsci.edu/hst/campaigns/fronteir-fields/publications}{http://www.stsci.edu/hst/campaigns/fronteir-fields/publications} for a more comprehensive list of references and publications.}

The Space Telescope Science Institute (STScI) contracted several lens modeling teams to compute lens models for each of the \hff\ clusters and make them available to the public. As part of this effort, the different teams collaboratively evaluated the robustness of the multiply-imaged systems that were found in these data. Other than agreeing on a uniform set of constraints, each team worked independently from the other teams. In that, the \hff\ initiative presents us with a unique opportunity to compare different algorithms that were used to model the same cosmic lenses, the various assumptions used by the lensing teams, learn about systematics in lens modeling, and the strengths of different algorithms. Comparison projects using simulated galaxy clusters \citep{Meneghetti:17,Acebron:17} and \hff\ clusters \citep{Priewe:17} investigate the performance of the algorithms employed by the lensing modelers. Other projects contrasted lens model predictions with observations from strongly lensed supernovae \citep{Rodney:15,Kelly:16,Rodney:16,Treu:16}.

Since the beginning of the \hff\ program several versions of the lens models have been computed by each team. With each new version the quantity of constraints has increased due to the deep \hst\ observations allowing for identification of faint multiply-imaged systems of background sources. In addition, the quality of the constraints has improved with spectroscopic follow-up providing confirmation of the arcs, and importantly, spectroscopic redshifts (e.g., \citealt{Jauzac:14,Richard:14,Treu:15,Caminha:17,Karman:17,Mahler:18}). 

The goal of this paper is to examine the predictive power of strong lensing models. We evaluate the ten public \hff\ version 3 lens models for \cname\  taking advantage of newly measured spectroscopic redshifts \citep{Caminha:17} that were not available at the time the lens models were computed. We investigate how the different algorithms perform in regions of the field of view where there were no constraints when the lens models were computed. We also check for any dependence of the root-mean-square (RMS) scatter of the systems on the source spectroscopic redshifts. With the new data we can explore the predictive power of the lens models through the source plane scatter and image plane projections.
This is important for future predictions made using these models including the location of new multiple images of known or newly-discovered lensed sources.

This paper is organized as follows. The importance of source redshifts in strong gravitational lens models, \cname, and the lens models available from \hff\ are described in \S \ref{sec:background}. In \S\ref{sec:stronglensing}, we describe the lensing equation and the complexities of using the image plane scatter. Next, we discuss the methodology and results from the source plane and image plane projection analysis. \S\ref{sec:v4} examines the V4 models for \cname, which used the newly measured redshifts as constraints. Finally, a summary and conclusions can be found in \S\ref{sec:conclusion}. We assume a flat cosmology with $\Omega_M = 0.3$, $\Omega_\Lambda = 0.7$, and $H_0 = 70 \mathrm{km s} ^{-1} \mathrm{Mpc} ^{-1}$. 


\section{Background}
\label{sec:background}

\subsection{Strong Lensing Constraints}
\label{subsec:z_in_SL}

Strong lensing analysis relies on correct identification of sets of multiply-imaged lensed sources (``arcs''). Modelers use several selection criteria to identify multiple images, including morphology, color, redshift, and expected lensing geometry. Deep, high resolution observations increase the feasibility of finding arc candidates but come with an extensive cost in observational resources. These observations are crucial since both the accuracy and precision of the lens models improve with the use of more lensed galaxies, as has been shown by \citet{Richard:10b,Johnson:16}; and \citet{Mahler:18}.

The main constraints in strong lensing come from the observed positions of these images. Additionally, the redshift of the lensed source provides the distance (assuming a cosmology) giving us 3-dimensional information of the geometry of the system (\autoref{eq:lenseq}). Ideally, all of the multiply-imaged systems that are used to constrain the model would have spectroscopic redshifts. This is not typically the case, due to the high observational cost of spectroscopic follow-up. As the community devotes more resources to spectroscopic follow-up observations, the data become more complete. In the cases that spectroscopic redshifts are not available, lens modeling algorithms can treat redshifts as free parameters, increasing the flexibility of the models and thus increasing their uncertainty.

In the absence of spectroscopic redshifts, photometric redshifts can be used in the modeling process. Photometric redshifts are prone to systematic uncertainties, which can be reduced with information from multiple band-passes and deeper data, although catastrophic failures are possible. Nevertheless, photometric measurements can be implemented in the modeling process by using their posterior probability distribution as a prior for the free redshift parameter of the lensed source.

An investigation of the systematics in strong gravitational lens modeling exploring the importance of using spectroscopic redshifts was done by \citet{Johnson:16}, who found that the RMS scatter in the image plane improves when increasing the number of spectroscopic systems used in computing the lens models. \citet{Cerny:17} investigated methods of treating the redshifts of multiple images in various galaxy clusters and the effects on the magnifications computed from the lens models. The systematic uncertainties due to a small fraction of spectroscopic redshifts compared to a large fraction was analyzed in the case of Abell 2744 \citep{Mahler:18}, which was observed as part of the \hff\ program. The systematic effect on the retrieval of cosmological parameters from strong lensing models due to the different density profiles used by parametric lens models and the effect of redshifts were explored by \citet{Acebron:17}, using simulated galaxy clusters \citep{Meneghetti:17}.

In this work we take advantage of the factor of two increase in the number of lensing constraints with spectroscopically confirmed redshifts, to evaluate lens models of the \hff\ cluster \cname.


\subsection{The Predictive Power of Lensing Models in Clusters of Galaxies}
\label{subsec:predictivepower}

All strong lens models rely on lensing observables as constraints. A variety of predictions can be made from the computed models, including the locations of new multiple images of known or newly-discovered lensed source. The use of lens models to find background source projections is standard practice by lens modelers as well as the rest of the community, and is routinely used to discover new images of lensed background sources, which in turn could help provide new constraints for the lens models.

With a well-constrained lens model, the angular separation between images of the same source can be used to estimate a ``geometric'' redshifts in the absence of spectroscopic redshift (e.g., \citealt{Zitrin:14,Chan:17}). In addition, model-predicted redshifts can be compared to independent photometric redshift estimates, and help discriminate between degenerate photometric redshifts (e.g., \citealt{Lam:14,Diego:15b,Cerny:17}).

Time delay can also be predicted by lens models. Since the travel time of light depends on its path between the source and the observer, different images of the same source are not contemporaneous. The arrival time difference, or time delay, between the multiple images of the same source, depends on the lensing potential and can be directly derived from a lens model. 
If the source is transient or variable in nature, the time delay can be observationally measured and compared to lens model predictions, e.g., Supernovae \citep{Kelly:14,Oguri:15,Sharon:15a,Diego:16b,Jauzac:16,Kelly:16,Treu:16,Rodney:18} and quasars lensed by clusters \citep{Inada:06,Fohlmeister:07,Fohlmeister:13,Dahle:13,Sharon:17}.

Finally, the lensing magnification is seldom a direct observable. In most cases, one can predict and measure the {\it relative} lensing magnification between images of the same source. The only known exception thus far is an observed standard candle such as supernova Type~Ia, whose peak magnitude can be known through calibration and thus its absolute magnification can be compared to model predictions \citep{Rodney:15}. 

In this paper, we focus on the ability of lens models to properly predict the source location of lensed images. For assessment of the predicting power of magnification or time delay, the reader is referred to the publications listed above.


\subsection{\cname}
\label{subsec:\cname}

\cname\  is a massive cluster at a redshift $z=0.397$ identified in the MAssive Cluster Survey (MACS \citealt{Ebeling:07, Ebeling:14}). It was suggested that it is a merging cluster \citep{Mann:12} evident by the double-peak profile of its bolometric X-ray luminosity. The Cluster Lensing and Supernova survey with {\it{Hubble}} (CLASH) has studied this cluster to constrain its mass distribution \citep{Postman:12}. Using the CLASH \hst\ images, \citet{Zitrin:13} identified 70 multiple images of 23 background sources, which were used in the strong lensing analysis. This cluster was previously observed by {\it{Spitzer}} IRAC (PI R. Bouwens), {\it{Chandra Space Observatory}}, Jansky Very Large Array (JVLA), Giant Meterwave Radio Telescope (GMRT) \citep{Ogrean:15}, and Very Large Telescope (VLT) spectroscopy (e.g., \citealt{Grillo:15,Balestra:16}). As part of the \hff\ program, \cname\  was observed again by \hst\ in early 2014 and late 2014. Some of the most recent work exploring \cname\  include joint x-ray and strong lensing analysis to investigate the dark matter distribution in the core of the galaxy cluster \citep{Bonamigo:17}. Line-of-sight structure is relevant when performing strong lensing analysis, since the lensing signal is sensitive to the total projected mass distribution \citep{Bayliss:14,D'Aloisio:14}. A study of line-of-sight structure effects on the strong lens modeling of \cname\  was presented by \citet{Chirivi:17}. \citet{Natarajan:17} studied the dark matter substructure of \cname\  by comparing lensing inferred substructure to that of simulations.

\begin{figure}
\includegraphics[width=0.5\textwidth]{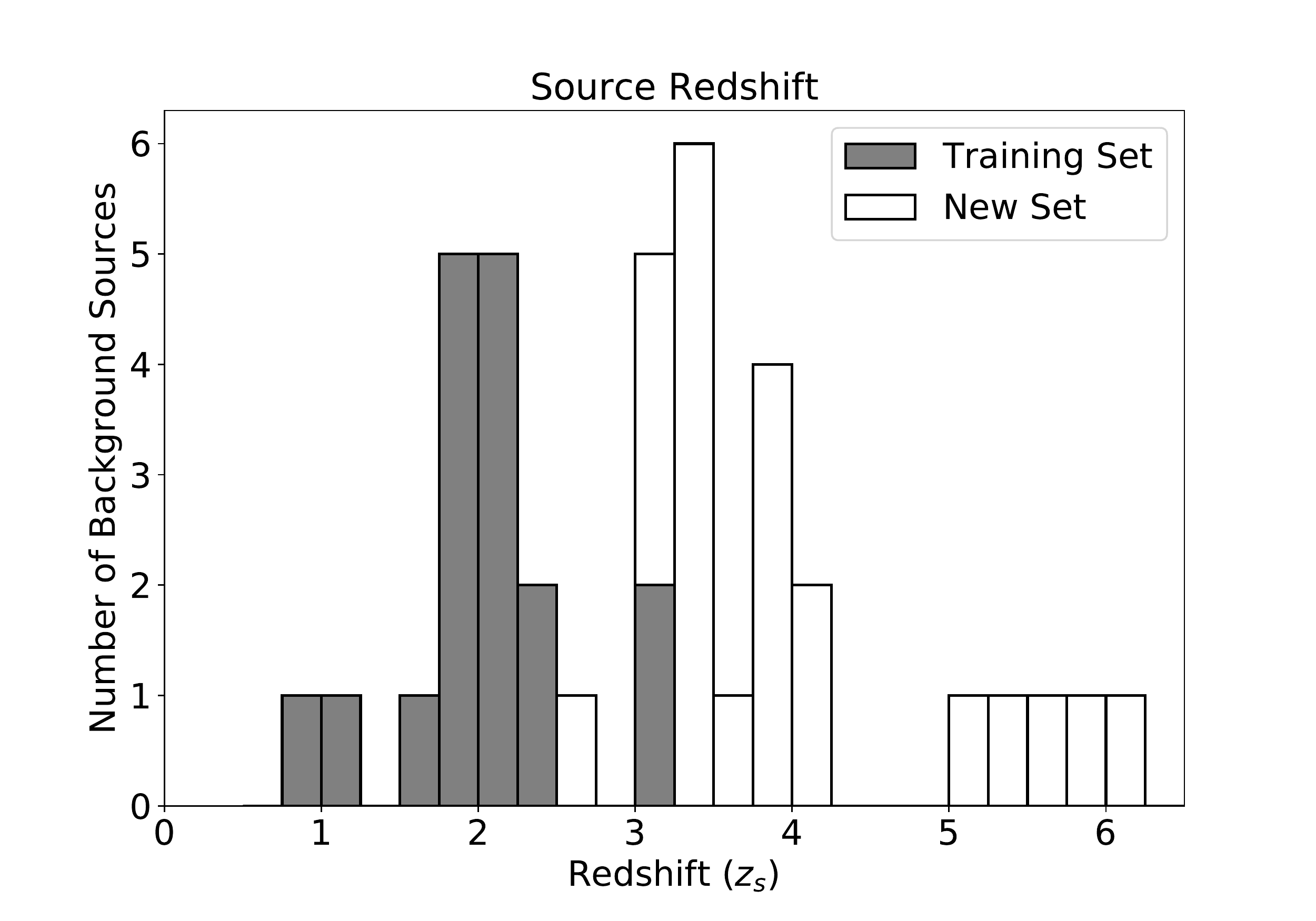}
\caption{Redshift distribution of the background sources that are lensed by the galaxy cluster \cname\  are plotted. The gray area represents the training set (data available to the lensing groups when computing the models in 2015 from \citealt{Jauzac:14,Richard:14,Treu:15}) and the white is the new set (data available from the resent follow-up survey by \citet{Caminha:17} that did not have any previous archival spectroscopic redshifts). From the distribution, it is clear that the new arcs tend to have a higher redshift than those arcs in the training set. This color convention is going to be carried through the rest of of the paper.}
\label{fig:zs}
\end{figure}

\begin{figure*}
\center
\includegraphics[width=0.7\textwidth]{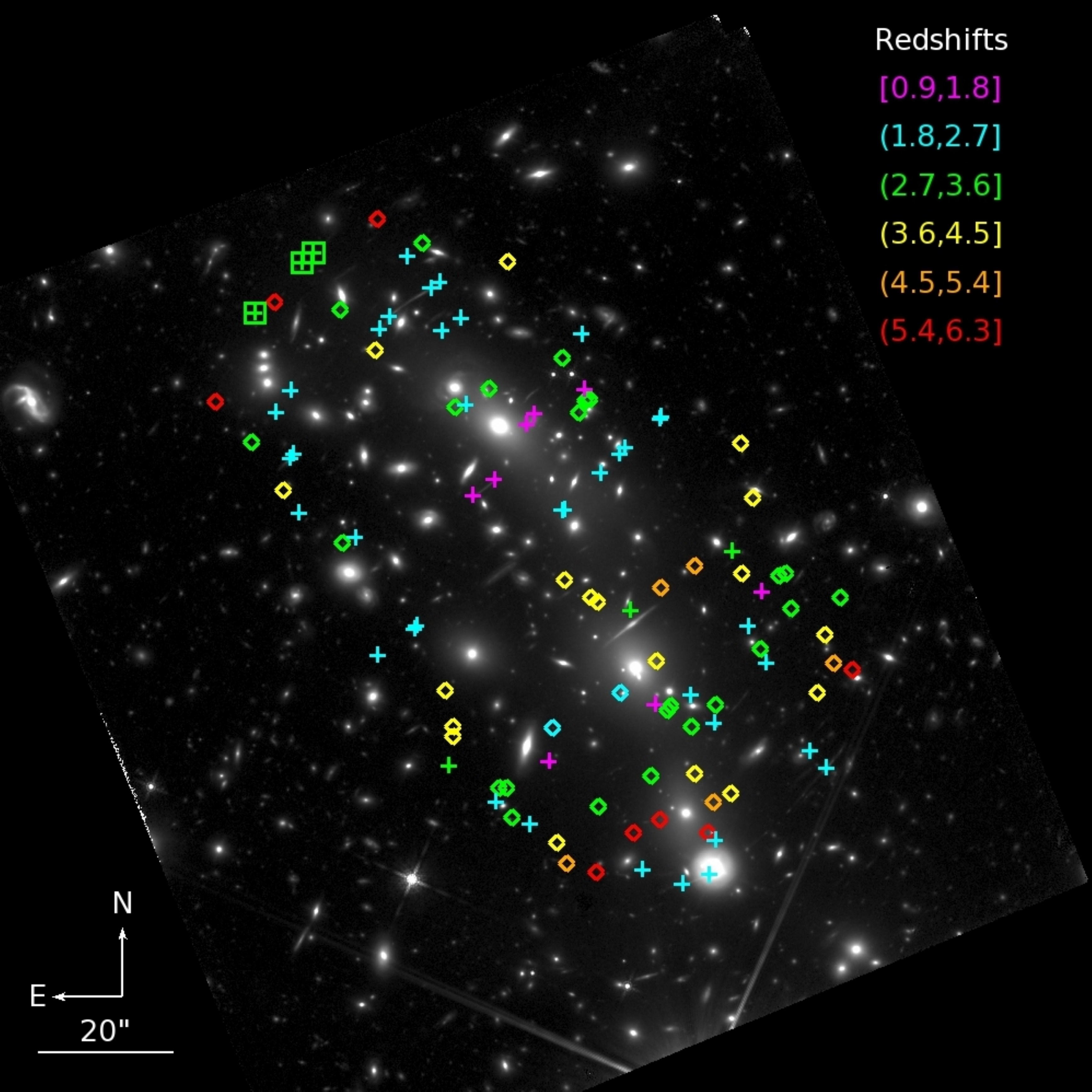}
\caption{\hst\ optical F160W image of \cname\  where the colored marks represent the position of the multiply-images, and the color indicates the background-source-measured spectroscopic redshift \citep{Jauzac:14,Richard:14,Treu:15,Caminha:17}. The plus symbols represent the training set while the diamonds indicate the new lensed systems. The boxes indicate lensed system 26, which had a spectroscopic redshift update from $z_{26} = 2.1851$ to $z_{26} = 3.2355$.}
\label{fig:zs_loc}
\end{figure*}

With the deep \hst\ imaging, hundreds of multiple images have been identified and spectroscopic redshifts measured in follow-up observations. In 2015, seven independent lensing teams computed lens models for \cname\  with the positional and redshift constraints available at that time. This included 17 sources with confirmed $z_{spec}$ \citep{Jauzac:14,Richard:14,Treu:15}. In this analysis, we refer to the set of lensed images with spectroscopic redshifts available when the lens models were computed as the ``training set''. In 2016, a new follow-up survey by \citet{Caminha:17} increased the number of spectroscopic redshifts to 39 sources and updated some of the previously found redshifts. The lensed images (arcs) with new spectroscopic redshifts are referred to as ``new set'' in our analysis. \autoref{tab:sys_img_data} has a summary of the number of background sources and the multiple images with spectroscopic redshifts for \cname. The distribution of the source spectroscopic redshifts ($z_s$) is shown in \autoref{fig:zs}. We note that most of the new arcs have a higher redshift compared to the training set. Out of the spectroscopic redshifts that were updated, only system $26$ had a drastic change in $z_{spec}$ from $z_{26} = 2.1851 \text{ to } z_{26} = 3.2355$ (see \citealt{Caminha:17}). In \autoref{fig:zs_loc}, we over-plot the positions of the arcs on a WFC/IR (F160W) image of \cname. The colors represent the source spectroscopic redshift, the symbols indicate if the lensed system is part of the training (plus) or new (diamond) set, and the box identifies the location of system 26.

\capstartfalse
\begin{deluxetable}{ccc}[h]
\tablecolumns{3}
\tablewidth{.8\columnwidth}
\tablecaption{Number of arcs with $z_{spec}$ for \cname .}
\tablehead{\colhead{Sets} & \colhead{$N_{sys}$} & \colhead{$N_{im}$}}
\startdata
Training & 17 & 51 \\ 
New & 22 & 57 \\ 
\hline \\
Total & 39 & 108 \\ 

\enddata
\tablenotetext{}{Number of multiple images ($N_{im}$) and background systems ($N_{sys}$) with spectroscopic redshifts ($z_{spec}$) used in our analysis.}
\label{tab:sys_img_data}
\end{deluxetable}
\capstarttrue

The independent lensing teams have ranked the lensed systems based on the confidence in identification of the multiple images, availability of spectroscopic redshift for the source, and lensing geometry. The arcs that had the highest reliability and measured spectroscopic redshifts were labeled ``GOLD''. Secure arcs with no available $z_{spec}$ were labeled ``SILVER''. Last, ``BRONZE'' were the multiply-imaged systems where there was neither a unanimous agreement on the reliability nor spectroscopic redshift confirmation. The lensing teams used these vetted selections to compute the lens models. We refer to these classifications in this paper.


\subsection{Lensing Models}
\label{subsec:lensing_models}

The lensing groups contracted by STScI computed independent lens models using a variety of algorithms, which can generally be separated into two categories: ``parametric'' and ``non-parametric'' (also called ``free-form''). Parametric algorithms use a parametrized density distribution for dark matter halos and cluster member galaxies, to determine the lensing potential and mass distribution of the cluster. Non-parametric or free-form algorithms make little to no assumption on the functional form of the density profile of the galaxy cluster. Hybrid models are some combination of the parametric and non-parametric algorithms. In addition to the different forms of modeling the dark matter halos, an assumption that light traces mass is used in some of the algorithms. 

The following is a brief overview of the models analyzed in this work. A summary of the lens models name, algorithm, and constraints used can be found in \autoref{tab:lens_models_data}. A more detailed description of the lens models can be found in \href{http://stsci.edu/hst/campaigns/fronteir-fields/Lensing-Models}{\hst\ Frontier Fields Lensing Models Web-page}.\footnote{See \href{http://stsci.edu/hst/campaigns/fronteir-fields/Lensing-Models}{http://stsci.edu/hst/campaigns/fronteir-fields/Lensing-Models} for the publicly available lens models.}

{\bf{CATS (C-V3, C-V3.1)}} collaboration \citep{Jauzac:12,Jauzac:14,Richard:14} and {\bf{Sharon/Johnson (JS-V3)}} team (hereafter JS; \citealt{Johnson:14}) use publicly available parametric lensing model software \texttt{LENSTOOL} \citep{Jullo:07,Jullo:09}. JS used only strong lensing while CATS used a combination of strong and weak lensing, for the reconstruction of the mass distribution of the galaxy cluster. The total mass distribution of the cluster is parametrized by a combination of cluster-scale halos and a contribution from individual cluster member galaxies, using scaling relations to observed galaxy properties. \texttt{LENSTOOL} uses a Bayesian Monte Carlo Markov Chain (MCMC) algorithm to explore the parameter space minimizing a $\chi^2$ to identify the best set of parameters. There are two models computed by CATS. Version 3 (C-V3) used only the ``GOLD'' and ``SILVER'' arcs, while version 3.1(C-V3.1) used ``GOLD'', ``SILVER'', and ``BRONZE''. The model by JS (JS-V3) only included the ``GOLD'' arcs.

{\bf{Glafic (G-V3)}} \citep{Oguri:10,Ishigaki:15,Kawamata:16} used a parametric algorithm where a large smooth elliptical Navarro-Frenk-White (NFW; \citealt{Navarro:97}) profile represents the cluster mass distribution while the individual galaxies are modeled using Pseudo-Jaffe ellipsoids \citep{Keeton:01}. To complement these two components, external and internal perturbations of the lensing potential are introduced and represented by a multipole Taylor expansion. The ``GOLD'' and ``SILVER'' arcs were used to constrain the lens model.

{\bf{Zitrin (Z-NFW, Z-LTM)}} models \citep{Zitrin:09,Zitrin:13} used two different methods of parametrization. The first is a NFW dark matter halo density to describe the cluster mass distribution. The second method, introduced by \citet{Broadhurst:05}, is called Light-Traces-Mass (LTM). LTM uses a parametric profile for the cluster member galaxies scaled to their luminosity, then the mass map is smoothed using a Gaussian kernel to represent the dark matter component of the cluster. Both Z-LTM and Z-NFW use the ``GOLD'' and ``SILVER'' arcs.

{\bf{Brada\v{c} (B-V3)}} \citep{Hoag:16} used Strong and Weak Lensing United (SWUnited \citealt{Bradac:06,Bradac:09}). This non-parametric model combines both strong and weak lensing analysis to constrain the gravitational potential through iterative minimization of $\chi^2$ of a non-regular adaptive grid. The ``GOLD'' arcs were used to compute the model.

{\bf{Williams (W-V3, W-V3.1)}} models \citep{Liesenborgs:06,Mohammed:14} use GRALE, a non-parametric method that uses a genetic algorithm to iteratively refine the mass distribution on a grid. The models only use the lensing observables as constraints with no other assumption on the mass density profile or cluster galaxies. This method is ideal for testing the assumption that light traces the mass of the cluster. The W-V3 model used the ``GOLD'' and ``SILVER'', while the model W-V3.1 used ``GOLD'', ``SILVER'', and ``BRONZE'' arcs.

{\bf{Diego (D-V3)\footnote{The publicly released version 3 lens model of Diego was not suitable for this study due to a scaling problem. We obtained the correct version of the files from Jose Maria Diego (private communication)}}} \citep{Diego:15a,Diego:16a} use WSLAP+ which is a free-form algorithm with the addition of a parametrized distribution for the mass contribution from the galaxy cluster members. The large scale dark matter distribution of the cluster is obtained using the non-parametric algorithm, by using a superposition of two-dimensional Gaussians in an adaptive grid. The compact mass contribution from the cluster members is tied to a light-to-mass ratio. The ``GOLD'' and ``SILVER'' arcs have been used as inputs to compute the lens model.

\capstartfalse
\begin{deluxetable*}{ccccc}[h]
\tablecolumns{5}
\tablewidth{2\columnwidth}
\tablecaption{\hff\ V3 lens models for \cname.}
\tablehead{\colhead{Lens Model} & \colhead{Algorithm Name} & \colhead{Algorithm} & \colhead{Constraints Used} & \colhead{Model Reference}}
\startdata
B-V3 & SWUnited & Non-Parametric & ``GOLD'' & \cite{Hoag:16} \\
C-V3 & LENSTOOL & Parametric & ``GOLD'', ``SILVER''& \cite{Jauzac:14,Richard:14} \\ 
C-V3.1 & LENSTOOL & Parametric & ``GOLD'', ``SILVER'', ``BRONZE'' & \cite{Jauzac:14,Richard:14}\\ 
D-V3 & WSLAP+ & Hybrid & ``GOLD'', ``SILVER'' & \cite{Diego:15a,Diego:16a}\\
G-V3 & GLAFIC & Parametric & ``GOLD'', ``SILVER'' & \cite{Kawamata:16}\\
JS-V3 & LENSTOOL & Parametric & ``GOLD'' & \cite{Johnson:14}\\
W-V3 & GRALE & Non-Parametric & ``GOLD'', ``SILVER'' & \cite{Mohammed:14}\\
W-V3.1 & GRALE & Non-Parametric & ``GOLD'', ``SILVER'', ``BRONZE'' & \cite{Mohammed:14}\\
Z-LTM & LTM & Parametric & ``GOLD'', ``SILVER'' & \cite{Broadhurst:05,Zitrin:13}\\
Z-NFW & PIEMD+eNFW & Parametric & ``GOLD'', ``SILVER''& \cite{Zitrin:09,Zitrin:13} \\
\enddata
\tablenotetext{}{This is a summary of the ten public version 3 models of \cname\ including the name, type of algorithm, and constraints used to compute the model. For definition of constraints, see \S\ref{subsec:\cname}}
\label{tab:lens_models_data}
\end{deluxetable*}
\capstarttrue


\section{Strong Lens Models Evaluation}
\label{sec:stronglensing}

The lensing equation provides a transformation between each image location in the image plane to that of its source in the source plane:

\begin{equation}
\begin{split}
	\vect{\beta}_s & = \vect{\theta}_i - \vect{\alpha} (\vect{\theta}_i), \\
    \vect{\alpha}(\vect{\theta}_i) & = \frac{\dls (z_l,z_s)}{\ds (z_s)} \vect{\hat{\alpha}} (\vect{\theta}_i),
\label{eq:lenseq}
\end{split}
\end{equation}

\noindent where $\vect{\beta_s}$ is the source plane position, $\vect{\theta_i}$ is the location in the image plane of image $i$ of source $s$, $\hat{\alpha}(\vect{\theta_i})$ is the deflection angle, $\dls(z_l,z_s)$ is the angular diameter distance between the lens and the source, and $\ds (z_s)$ is the angular diameter distance between the observer and the source.
Lens modeling algorithms find the distribution of projected mass density that minimizes the scatter between the observed and predicted images of all background sources. 
Ideally, the strong lensing analysis would be done in the image plane where the observables are found, typically, minimizing the equation \citep{Kneib:11}: 

\begin{equation}
	\chi_s^2 = \sum_{i=1}^{n_s} \frac{[\vect{\theta}_{obs}^i - \vect{\theta}_{model}^i]^2}{\delta_{si}^2},
\label{eq:imageplanechisquare}
\end{equation}

\noindent where $\vect{\theta}_{obs}^i$ and $\vect{\theta}_{model}^i$ are the observed and the predicted positions by the lens model of image $i$ of system $s$ respectively and $\delta_{si}^2$ is the error on the position of image $i$ in the lensed system $s$.

When assessing the image plane scatter of the different lens models for the analysis presented in this paper, we found that some models fail to predict all the multiple images for some of the systems (see \S\ref{sec:ippa}). In those cases, the scatter of the system would be driven by the largest angular separation between the observed images in the image plane, preventing a quantitatively meaningful assessment of the models.  

We overcome this issue by resorting to measuring the scatter in the source plane. Since the lens equation does not need to be inverted, there is always a prediction for $\vect{\beta_s}$. The source plane and image plane areas are related to each other through the lensing magnification, which can be different for each of the multiple images $\vect{\theta}_i$. We therefore average over the magnification-weighted contribution to the scatter of each image, as detailed below. 


\subsection{Source Plane Scatter per System}
\label{subsec:multiply-imaged}

To quantify the performance of the lensing models we use the magnification-weighted RMS scatter of the projected source positions in the source plane, with the goal of evaluating the predictive power of the models, and exploring its dependence on redshift and position in the image plane. 

The source-plane location $\beta_i$ of each image $i$ of source $s$ is calculated from the lensing equation (\autoref{eq:lenseq}). We use the deflection matrices ($\hat{\alpha}$) from the publicly released lens models, observed positions of the arcs ($\vect{\theta_i}$), and the source spectroscopic redshift data $z_s$ (see \autoref{tab:data}). The deflection maps are interpolated using a Bivariate Spline. The interpolation is critical to increasing the precision when determining the location in the source plane of the multiple images. 

Since the true position of the background source is non-observable, we compute the barycenter ($<\vect{\beta_s}>$) of the projected source positions \citep{Gade:10}, weighted by the magnification of each image:

\begin{equation}
	<\vect{\beta_s}> = \frac{\sum_{i=1}^{n_s} \mu_i \vect{\beta_i}}{\sum_{i=1}^{n_s} \mu_i},
\label{eq:barrycenter}
\end{equation}

\noindent where $i = 1,2,3,...,n_s$ is the index of each image of lensed system $s$, $n_s$ is the total number of multiple images in the system, and $\mu_i$ is the magnification of image $i$. Last the distances between the barycenter and source plane locations are measured using the Vincenty formulae \citep{Vincenty:75} to measure a RMS scatter in the source plane. The weighted source plane RMS scatter ($\sigma _s$) for each of the systems is computed as follows,

\begin{equation}
	\sigma _s = \left( \frac{1}{n_s^2} \sum_{j=1}^{n_s}  \frac{1}{\mu_j^2}\sum_{i=1}^{n_s} (\mu_i || \vect{\beta_i} - <\vect{\beta_s}> ||)^2 \right)^{1/2},
\label{eq:system_RMS}
\end{equation}

\noindent where the summation takes into account the different magnifications of the images. 

When calculating the average scatter of several systems, $\overline{\sigma}$, we use the formula: 
\begin{equation}
	\overline{\sigma} = \frac{1}{N} \sum _s ^N (\sigma_s),
\label{eq:model_RMS}
\end{equation}
\noindent where $s = 1,2,3,...,N$ is the index of the system, $N$ is the total number of systems, and $\sigma_s$ is the source plane scatter of system $s$ (\autoref{eq:system_RMS}). 

We note that even though all of the modelers used the same list of multiple images (as described in \S\ref{subsec:\cname}), the positions are slightly altered, for example, they may use a unique morphological feature or a bright clump in the arc as the positional constraint. The coordinates used in this paper are tabulated in \autoref{tab:data}. We tested whether the exact choice of coordinate within the arc affects the results of our analysis. We computed the percent difference in the source plane scatter between two sets of positional coordinates from two different lensing teams\footnote{The coordinates of the constraints are from the JS and Diego teams, private communication.}. We find that the source plane RMS scatter changes by less than $5\%$, between the coordinate sets. This test provides confidence that the results described in the following sections are robust to slight differences in the coordinates used by the different teams.


\subsubsection{General Results}
\begin{figure*}
\center
\includegraphics[width=1\textwidth]{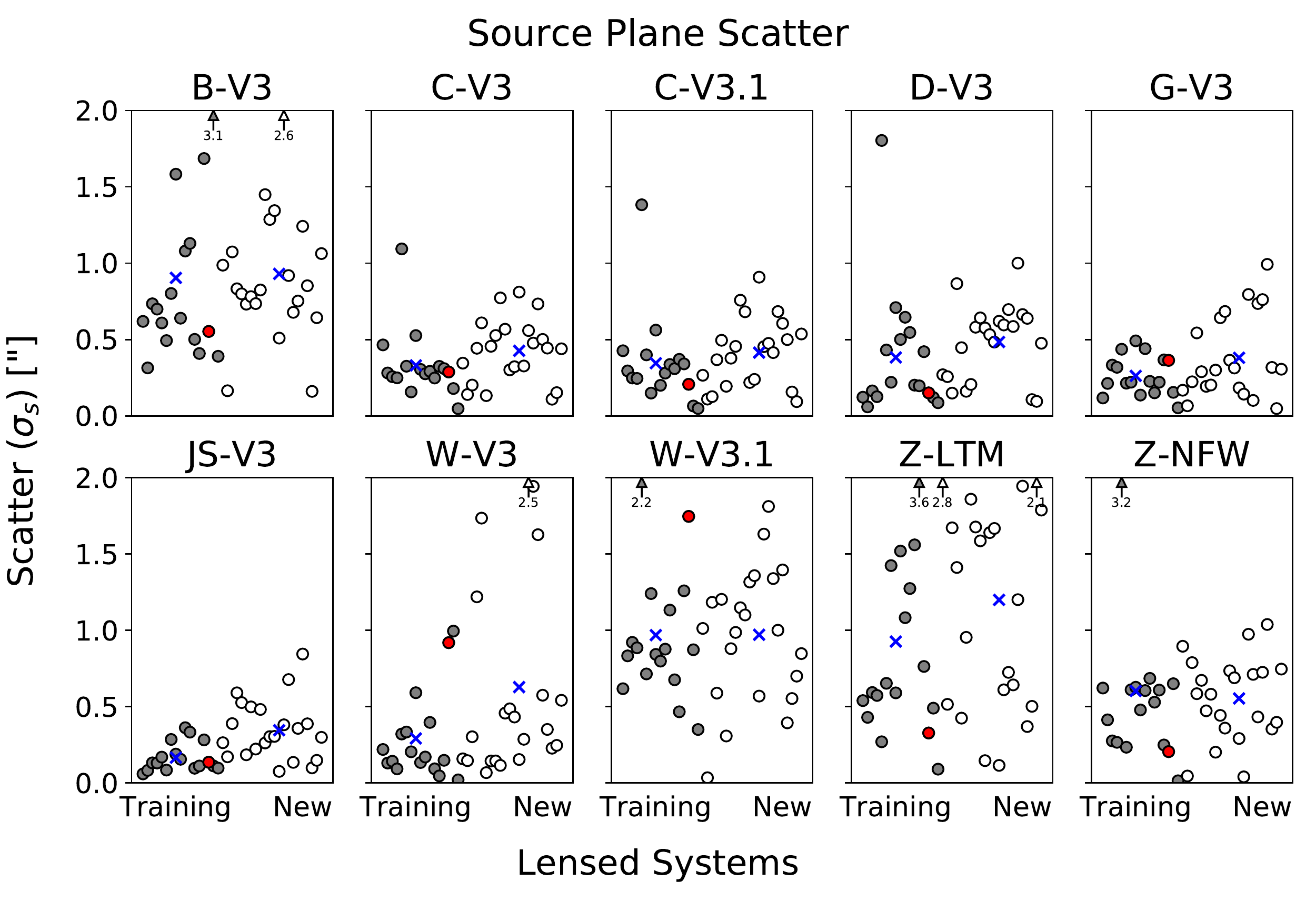}
\caption{The system RMS scatter for each lensed source ($\sigma_s$, \autoref{eq:system_RMS}) is plotted against the arc systems. Filled circles represents the systems that are part of the training set, and open circles represents systems that are part of the new set. The arrows at the top of the graph indicate data points that have a scatter $> 2\farcs 0$. The blue ``x'' represents the average scatter of the training and new systems respectively for each model ($\overline{\sigma}$, \autoref{eq:model_RMS}). The red dot marks system 26, whose redshift has been updated.}
\label{fig:system_sps}
\end{figure*}

\begin{figure*}
\center
\includegraphics[width=1\textwidth]{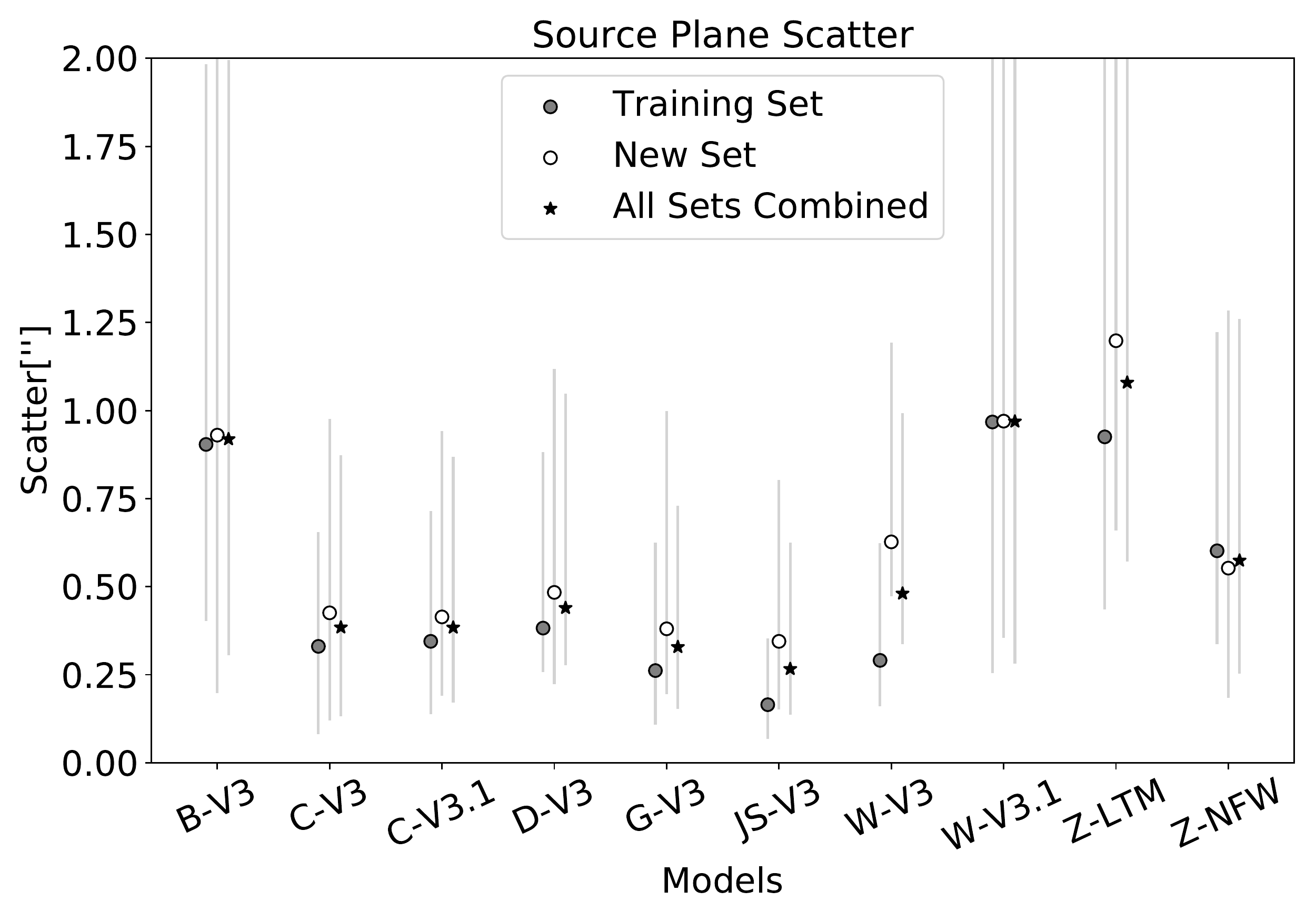}
\caption{The average model source plane scatter for the training, new, and combined sets for each of the models is plotted. The first and third quartiles are represented by the error bars. Most of the parametric models (C-V3, C-V3.1, G-V3, JS-V3, and Z-NFW) have similar RMS scatter with a combined RMS scatter $\overline{\sigma} < 0\farcs5$. The only parametric model with a high RMS scatter is Z-LTM. The Hybrid model, D-V3, has scatter which is similar to that of the parametric models. The non-parametric models, B-V3 and W-V3.1, have a higher RMS combined scatter compared to that of the parametric and hybrid models, while W-V3 has a low RMS combined scatter.}
\label{fig:models_sps}
\end{figure*}

In \autoref{fig:system_sps}, we plot the RMS scatter ($\sigma_s$) for each system for all the lens models, and in \autoref{fig:models_sps} we show the distributions (represented by $\overline{\sigma}$, first, and third quartiles) of the system RMS scatter of the training set, new set, and combined set, for a more direct comparison between the models. 

We find that models based on parametric or hybrid algorithms result in low RMS scatters (C-V3, C-V3.1, D-V3, G-V3, JS-V3, and Z-NFW), with combined average RMS scatter of $\overline{\sigma} < 0\farcs5$; The only parametric model which has a combined scatter of roughly $1\farcs 0$ is Z-LTM. On the other hand, most of the non-parametric algorithms result in higher RMS scatter (B-V3, W-V3.1, Z-LTM), with the exception of W-V3. 

Comparing the scatter of the training set to that of the new set, We find that with the exception of a few outliers, the training set systems tend to have smaller source plane scatter compared to the new systems. This is not surprising, since the training set was used to compute the models, meaning that the lens models are tuned to reproduce these multiple images.

\begin{figure*}
\center
\includegraphics[width=1\textwidth]{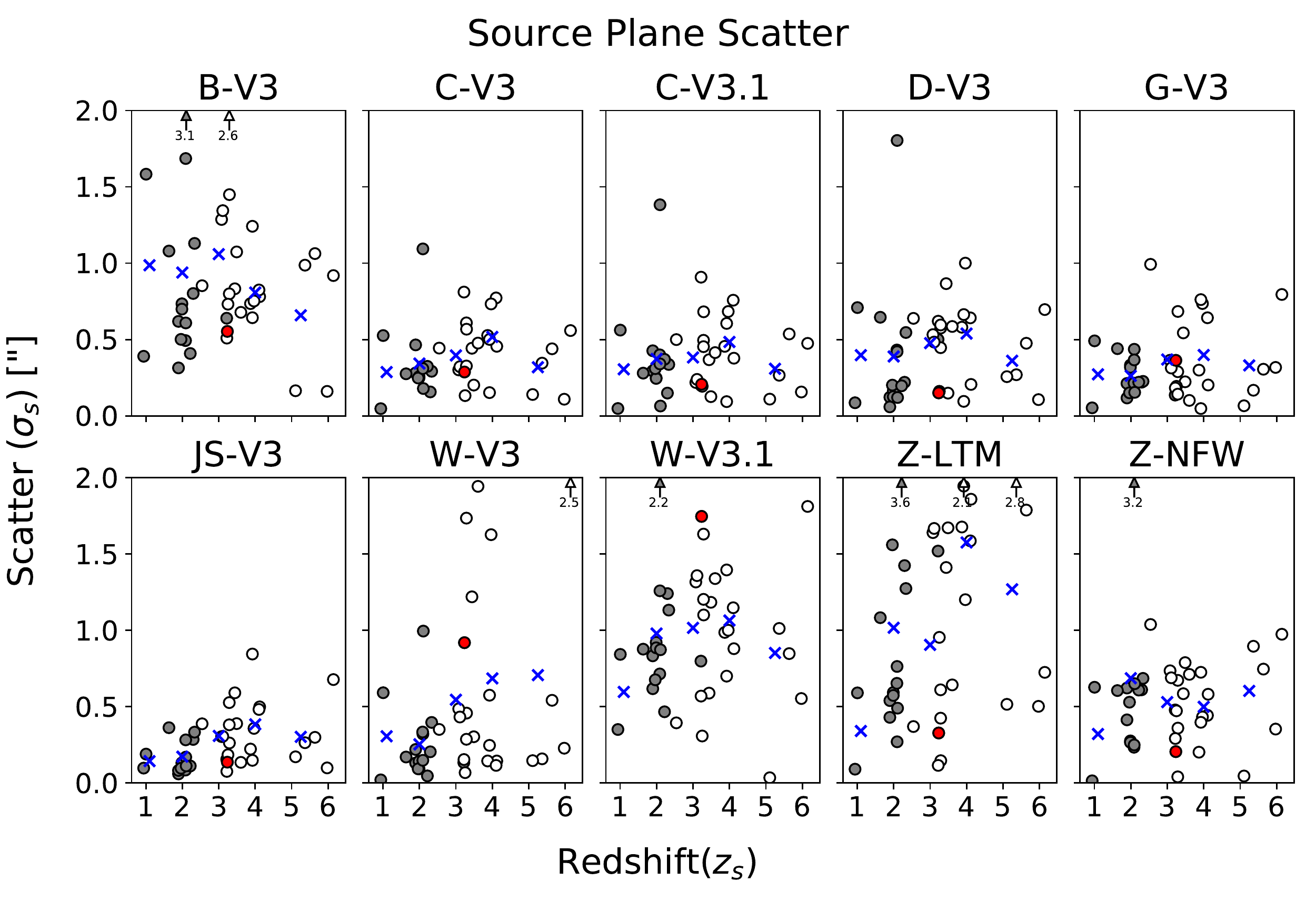}
\caption{The lensed system scatter ($\sigma_s$, \autoref{eq:system_RMS}) is plotted against the redshift of the system ($z_s$). The red circle shows the lensed system 26, plotted at its updated (correct) redshift. The blue ``x'' represents the average scatter for lensed systems in the following redshift bins: [0.9,1.5], (1.5,2.5], (2.5,3.5], (3.5,4.5], and (4.5,6.2].  We find no significant dependence on redshift within the training set (filled circles) and new sets (open circles). The apparent larger source scatter at higher redshifts is due to the fact that most of the sources in the new set are at higher redshifts as show in \autoref{fig:zs} and described in \S\ref{subsec:\cname}.}
\label{fig:redshift_sps}
\end{figure*}

We investigate whether the source plane RMS scatter ($\sigma_s$) increases or decreases with the system redshift ($z_s$) in \autoref{fig:redshift_sps}; we find no significant trend with redshift. The apparent larger source scatter at $z>2.5$ is due to the fact that most of the sources in the new set are at higher redshift.

Two teams computed additional models (V3.1) that include the ``BRONZE'' arcs, which allows us to assess their influence. When comparing C-V3 and C-V3.1, which use a parametric algorithm, we find that these two models have a similar RMS. This is not the case in models W-V3 and W-V3.1, which use a non-parametric method. Most of the systems in W-V3 have low scatter, except for a few outliers. In W-V3.1 the RMS scatter of all the systems is generally higher compared to that of W-V3. The difference between the V3 and the V3.1 models from both teams is the number and quality of constraints used when they were computed: V3 included ``GOLD'' and ``SILVER'' multiple images, while V3.1 used the ``GOLD'', ``SILVER'', and ``BRONZE'' arcs. This difference suggests that the GRALE algorithm produces better results when higher fraction of the lensing constraints are considered ``high quality''. Measured by the RMS diagnostic, it suggests that, at least for this particular cluster, adding low-confidence arcs that do not have spectroscopic redshifts degrades the lens model despite increasing the quantity of constraints. 

\subsubsection{Arc System 26, Source Plane Scatter}
\label{subsec:arc26_sps}

The latest spectroscopic follow-up by \citet{Caminha:17} confirmed most of the previously-known spectroscopic redshifts. One exception is the redshift of system 26, which was updated from $z_{26} = 2.1851$ \citep{Treu:15} to $z_{26} = 3.2355$ \citep{Caminha:17}. This change in redshift is of interest because it can be used to test how models perform when an erroneous spectroscopic redshift constraint is used to compute them.

\begin{figure}
\includegraphics[width=.5\textwidth]{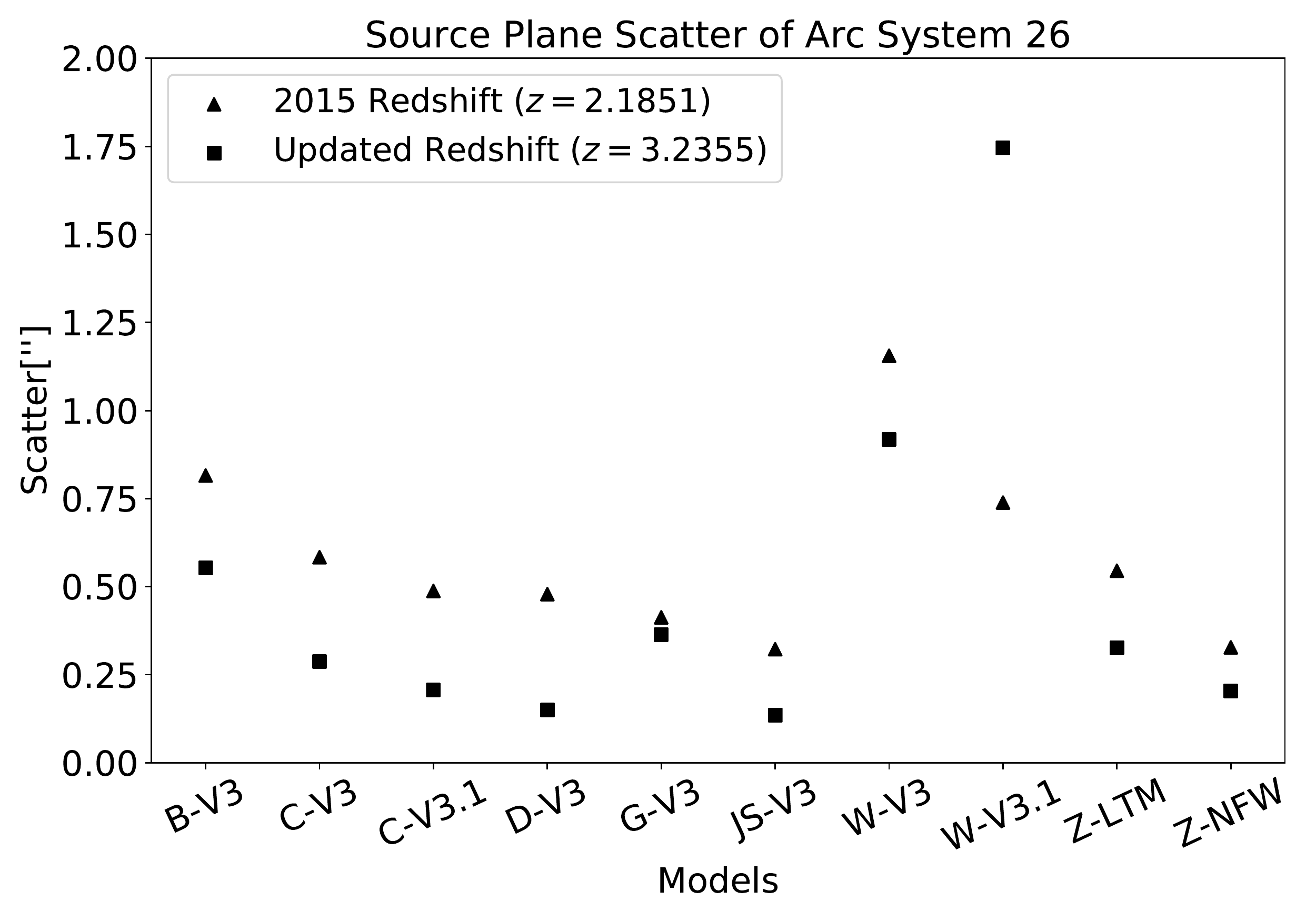}
\caption{The source plane scatter of the lensed system 26 ($\sigma_{26}$) is plotted for the different models. The redshift of this particular system was updated with the spectroscopic data from $z_{26} = 2.1851$ \citep{Treu:15} to $z_{26} = 3.2355$ \citep{Caminha:17}. For most of the lens models the RMS scatter is lower with the updated spectroscopic redshift, except for W-V3.1 in which the scatter increased by $\sim 1\farcs0$.}
\label{fig:sps_arc26}
\end{figure}

In \autoref{fig:sps_arc26}, we plot the RMS scatter of system 26 ($\sigma_{26}$) for both of the $z_s$ values for each model. The difference in source plane system scatter is only due to the change in the spectroscopic redshift. In most of the models, the RMS scatter of system 26 is about $0\farcs25$ lower with the new $z_{spec}$, meaning that system 26 is better reproduced by the lens models when the corrected spectroscopic redshift of $z_{26} = 3.2355$ is used. The only model in which the RMS scatter increased was W-V3.1, with a change of $\sim 1\farcs 0$, however this model had generally higher source plane scatter (\autoref{fig:system_sps}) and therefore this deviation may not be informative. A visual inspection of the positions of the images of system 26 relative to the magnification maps for the two redshifts indicate that in both cases the lensing configuration of this system is not reproduced well by this model. These changes show that in the case like the \hff\ clusters which have dozens of systems with spectroscopic redshifts, the effects of one erroneous redshift can be averaged out when computing the model average RMS scatter.


\subsubsection{Source Plane Scatter Dependence on Image Location}
\label{subsec:location_sps}

\begin{figure*}
\center
\includegraphics[width=1\textwidth]{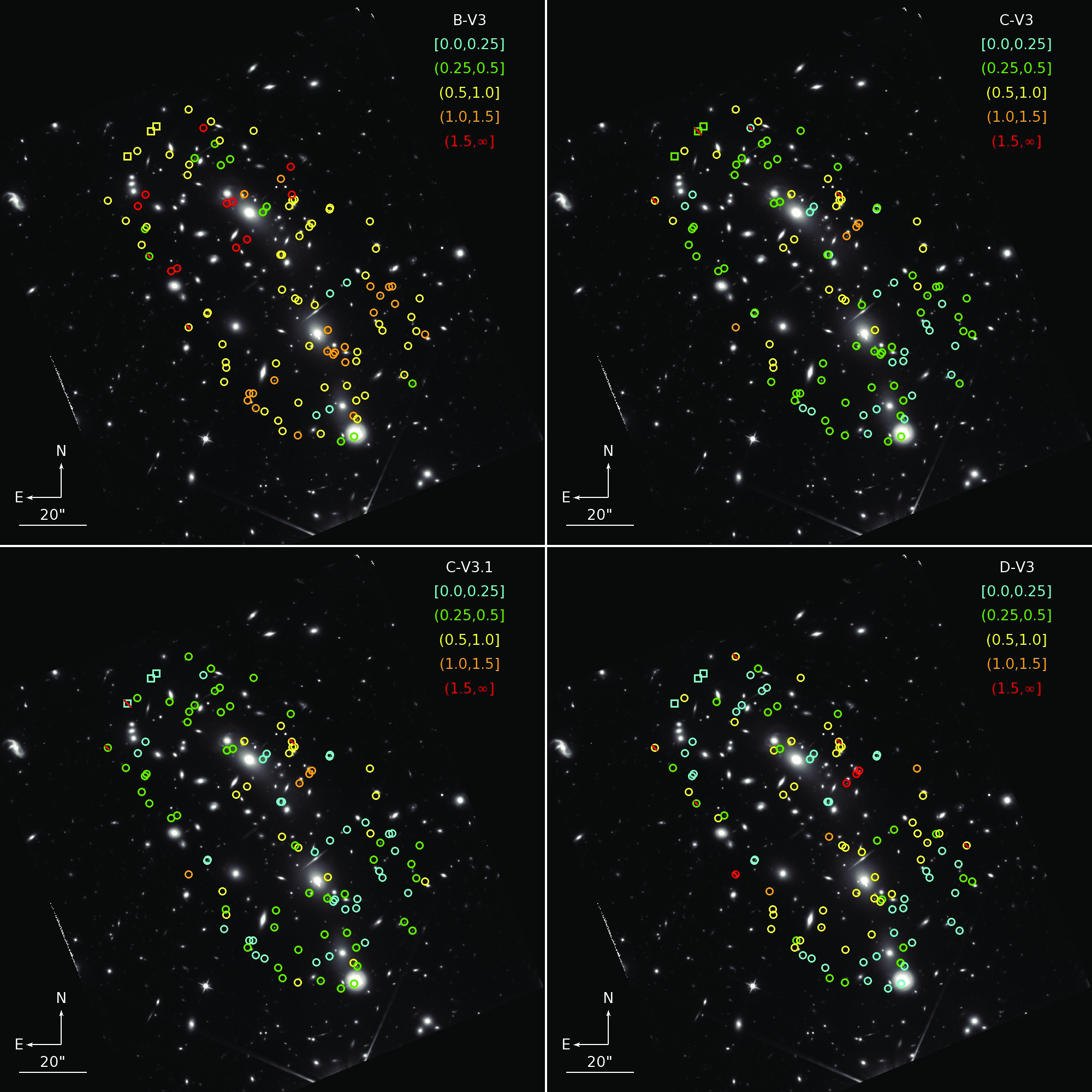}
\caption{\hst\ optical F160W image of \cname\  where the positions of the multiple images are indicated by the circles, while the color represents the source plane scatter of the lensed systems. We find models were the RMS scatter is uniform across the image plane, while some others have regions of relative high and low RMS scatter. The square symbols represent the location of the lensed system 26, and the square or circle with a line across the center, e.g., $\oslash$, represent ``failed'' arcs (see Section~\ref{sec:ippa}).}
\label{fig:loc_sps}
\end{figure*}

\begin{figure*}
\center
\includegraphics[width=1\textwidth]{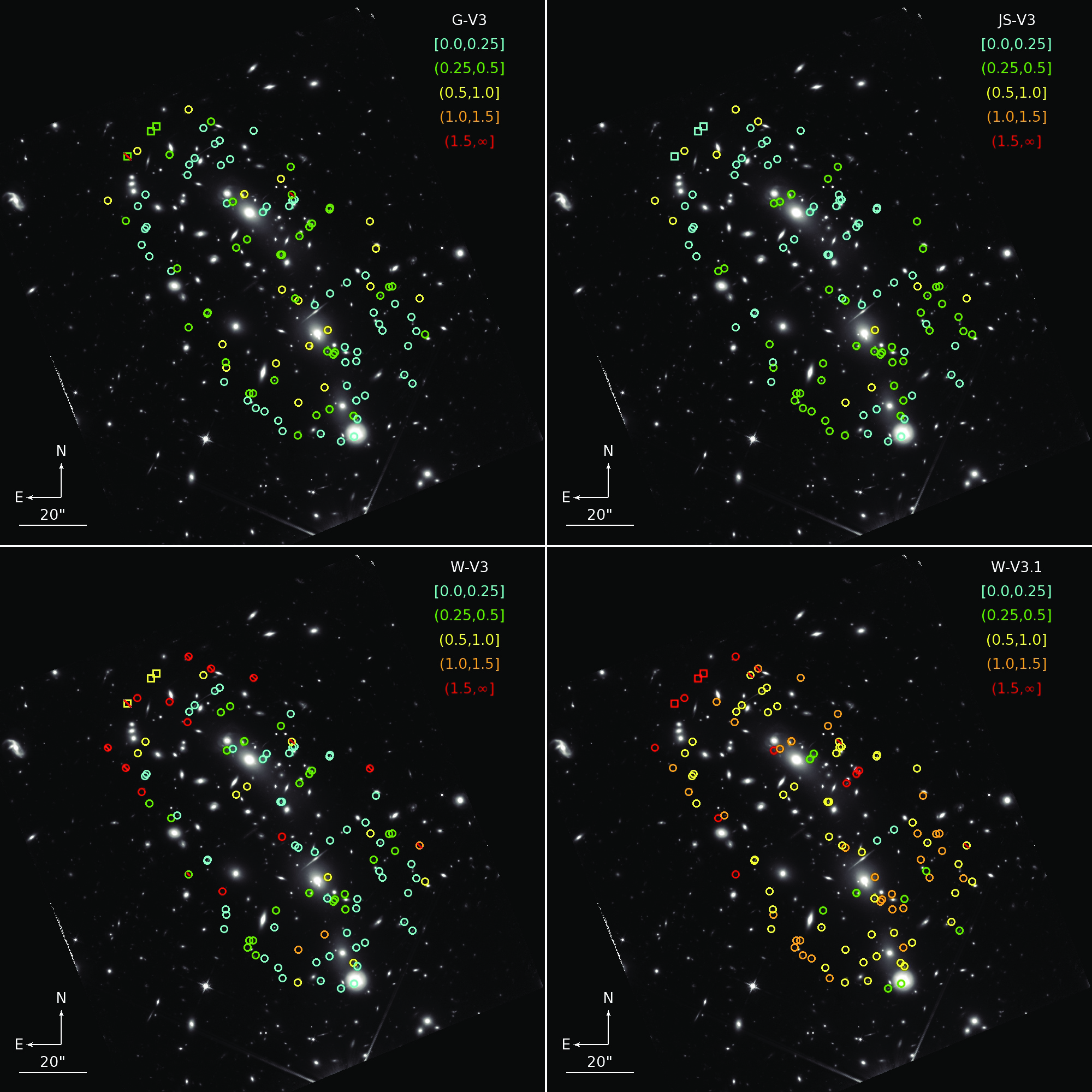}
\caption{Same as \autoref{fig:loc_sps}.}
\label{fig:loc_sps2}
\end{figure*}

\begin{figure*}
\center
\includegraphics[width=1\textwidth]{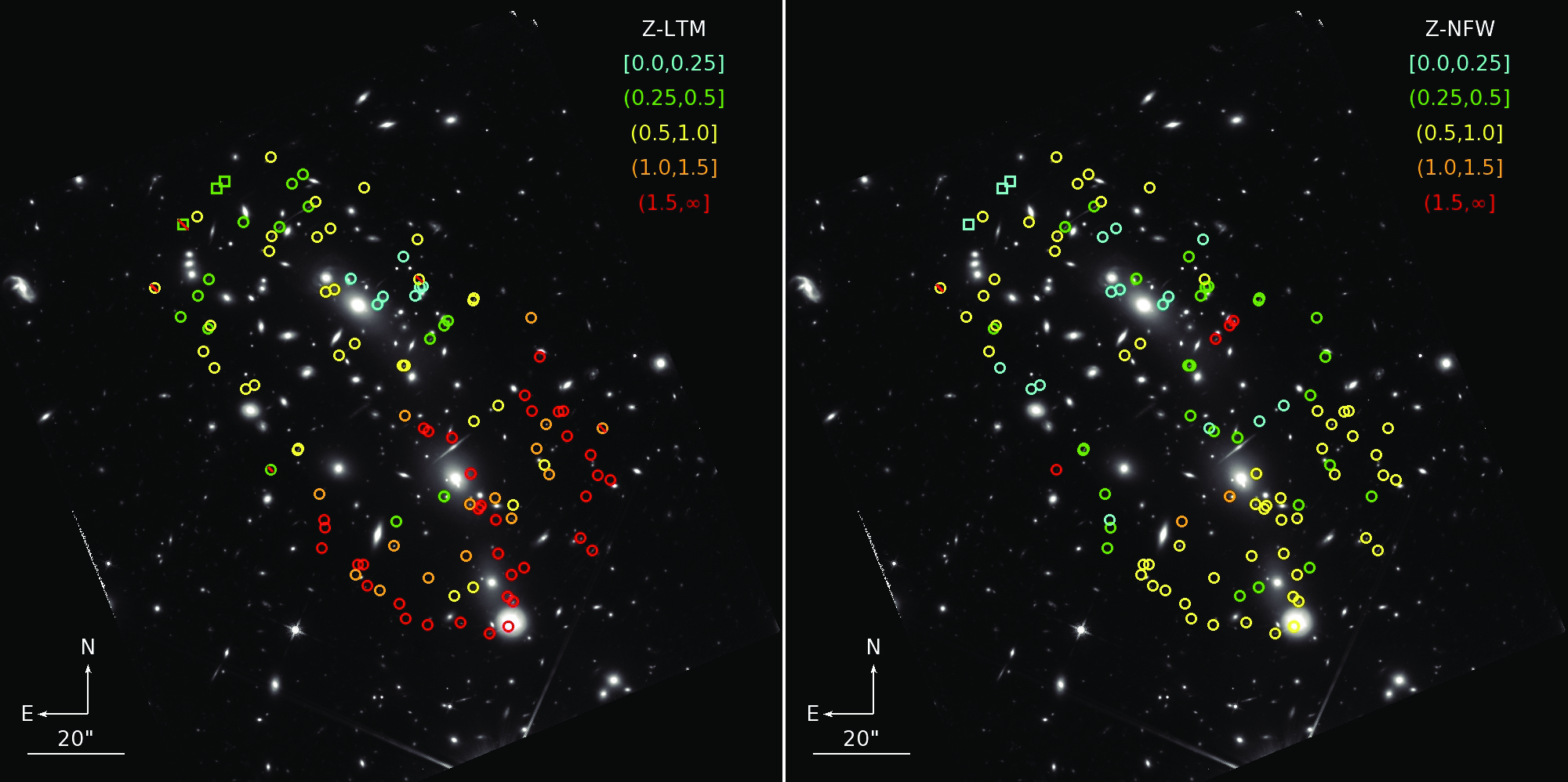}
\caption{Same as \autoref{fig:loc_sps}.}
\label{fig:loc_sps3}
\end{figure*}

Exploring the RMS scatter of systems with respect to their location in the image plane provides us with information related to regions in the image plane where a model performs well and regions where a model does not. In \autoref{fig:loc_sps}, \ref{fig:loc_sps2} and \ref{fig:loc_sps3}, we over-plot the positions of the multiple images on a WFC/IR (F160W) image of \cname. The source plane system RMS scatter is color coded as shown in the legend.

For some of the models (C-V3, C-V3.1, D-V3, G-V3, and JS-V3) the source plane RMS scatter is uniformly low across the image plane, while others show large variation from region to region in the image plane.  In addition, the similarity between C-V3 and C-V3.1, implies that when these parametric models already have a large number of constraints with spectroscopic redshifts, they do not benefit from adding low-score constraints for which no spectroscopic data are available.

The distribution of the system scatter in B-V3 is roughly uniform throughout the image plane with a slightly higher RMS scatter in the south-west region of the field of view. 

Both Zitrin models, Z-LTM and Z-NFW, have similar behavior across the field of view, with lower RMS values in the north-east and relatively high source plane RMS scatter in the south-west. However, Z-LTM has higher RMS scatter overall, with RMS higher than $1\farcs5$ in the south-west. Most of the systems in Z-NFW have an RMS scatter lower than $1\farcs0$ (\autoref{fig:loc_sps3}).

Last, we observe drastic difference between W-V3 and W-V3.1. In W-V3, most of the high-scatter systems are found in the north-east region of the image plane. In contrast, W-V3.1  has a more uniform RMS scatter through the image plane but is generally higher everywhere compared to W-V3. From this we can conclude that this model is strongly affected by the quality of the constraints --  adding the ``BRONZE'' images (low level of agreement on the robustness of the arc, and no spectroscopic redshifts) increased the RMS scatter in general across the entire image plane.   


\subsection{Evaluating the Models Predictive Power}
\label{subsec:predictive_power}

To evaluate the average predictive power of the lens models, we use the distribution of the systems source plane RMS scatter (\autoref{fig:system_sps}). Part of the predictive power is the ability of a model to accurately predict the position of new multiple images, with similar accuracy to that of the set of images that were used as constraints. To do so, we compare the RMS scatter of the new systems to that of the training set. 

\begin{figure*}
\center
\includegraphics[width=1\textwidth]{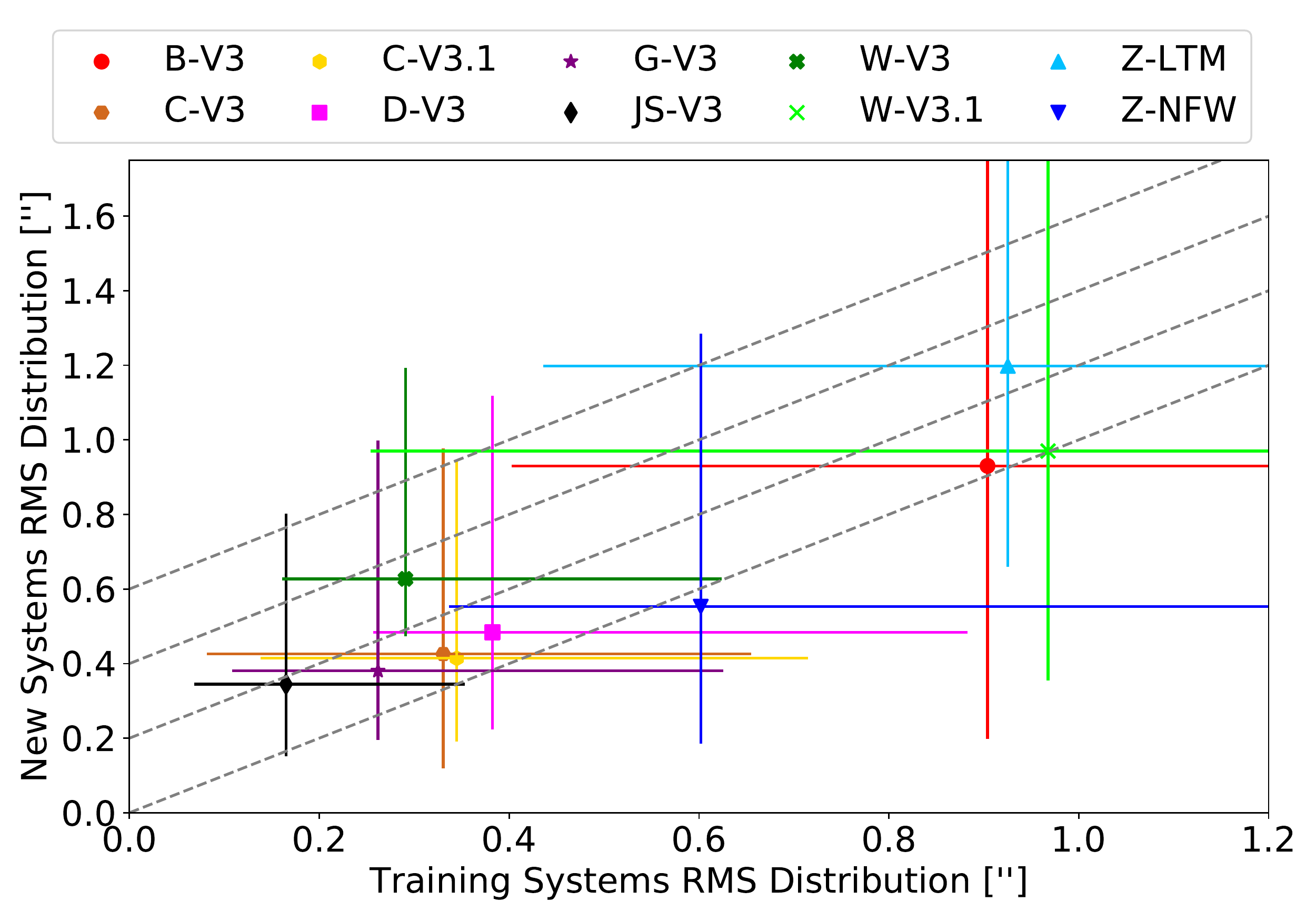}
\caption{To evaluate the predictive power of the lens models we plot the distribution of the training and new system RMS scatter. The error bars represent the first and third quartiles, and the symbols represent the arithmetic mean (\autoref{eq:model_RMS}) of the distribution. The dashed gray lines are there to guide the eye, they have a slope of one and increase by $0\farcs2$. The models with best predictive power are those which are closest to the $1:1$ line (lowest gray dash line) and have the lowest distribution for the training set.}
\label{fig:delta_models_sps}
\end{figure*}

In \autoref{fig:delta_models_sps}, we plot the distributions of the training set and new systems. We have added dashed gray lines to guide the eye, they have a slope of one and increase by $0\farcs2$. The models with the best predictive power are those closest to the $1:1$ line (lowest gray line) and low average training set RMS; these models, C-V3, C-V3.1, D-V3, G-V3, JS-V3, and Z-NFW, maintain similarly low, or only slightly higher average RMS scatter when confronted with lensing evidence that were not used to constrain the model. W-V3 has a low training set mean RMS, although when compared to that of the new set, we find it is further away from the self-similar line meaning that the model did not perform as well with the new images. In the case were  the models have a high training set arithmetic mean and are close to the $1:1$ line include W-V3.1 and B-V3. These models perform similarly between the two sets of images with relative high RMS scatter compared to the rest of the models. Last, Z-LTM has high training set RMS distribution and is far from the self-similar line, meaning that the model already has a high RMS scatter and the scatter increases even more when confronted with lensing evidence that were not used to constrain the model.  


\subsection{Image Plane Projection Analysis}
\label{sec:ippa}

In this section, we test the ability of the models to recover the observed lensing configuration in the image plane. 
We do this by computing the predicted source position of each image of each system using \autoref{eq:lenseq}, then lens it back to all of its predicted positions in the image plane. A correct prediction of the lensing configuration will result in a number of images in the image plane that is at least the same as the number of observed images in each system. It may sometimes predict additional images that may be unobservable, e.g., due to low magnification or obscuration by foreground sources.

To allow for conservative positional uncertainties, we set a circular area with radius of $0\farcs1$ around the location of the source plane position. We then ray trace the area inside the circle to the image plane to predict its multiple images. The predicted images are identified using the Hoshen-Kopelman \citep{Hoshen:76} cluster labeling algorithm. The original image should always be reconstructed. If more predictions appear, it means that the model predicts multiple images, corresponding to the circle that we had set in the source plane. In the case where no additional predicted images are found in the image plane, then the model doesn't predict multiple images from the circle in the source plane. We refer to these multiple images as failed arcs, because they fail to create some or all of the multiple projections.

The process of finding the other image plane projections is then repeated for each one of the multiple images. An arc fails to produce counter images if its source circle falls entirely outside the caustic (the source plane projection of the critical curve, representing loci of high $\mu$ in the image plane) for its redshift, in which case the model cannot predict the correct number of multiple projections in the image plane. 

\begin{figure}
\center
\includegraphics[width=.5\textwidth]{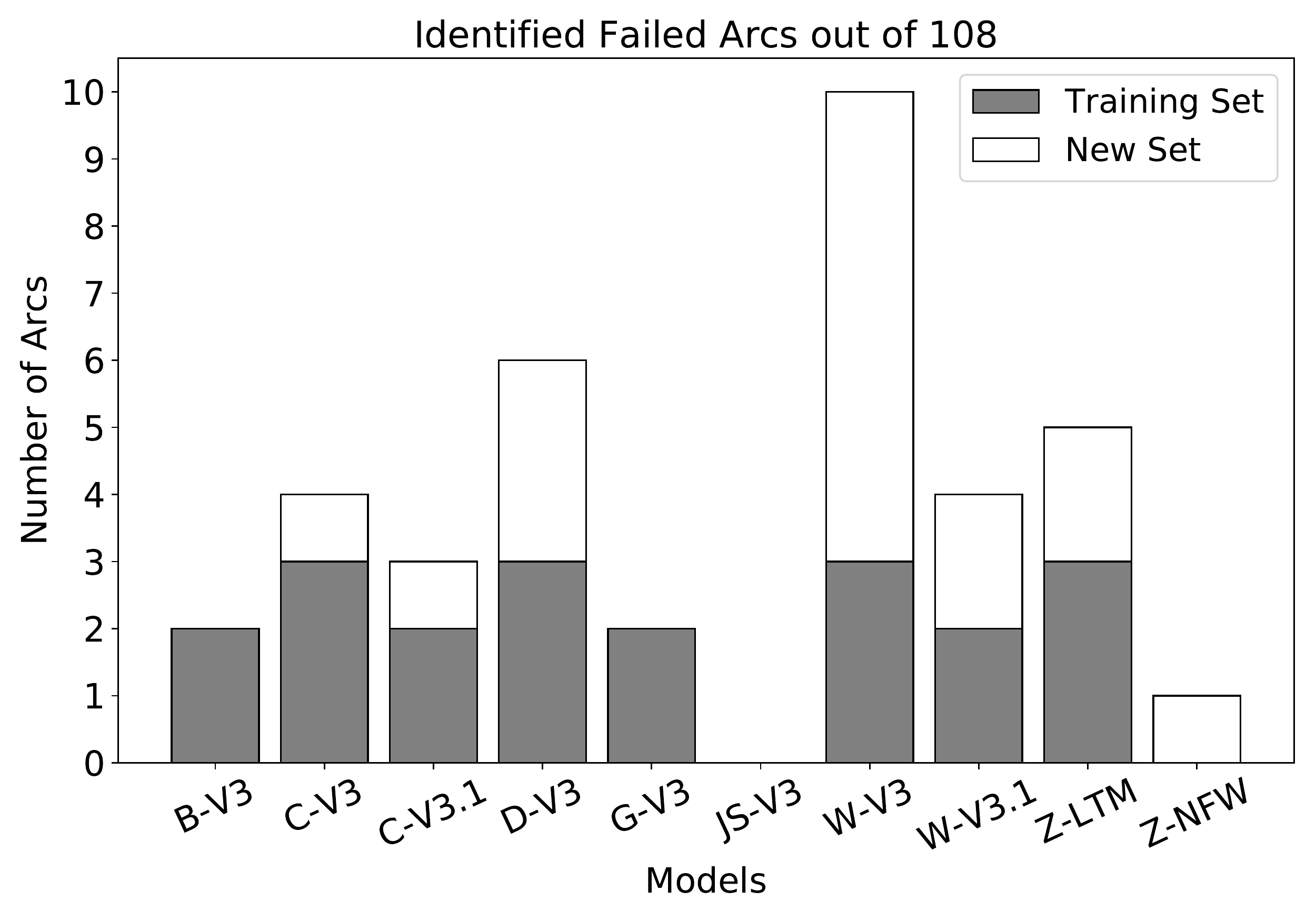}
\caption{Number of the arcs, out of the 108 total, that were labeled as failed arcs as follows: we compute the source plane position of the observed arc, then lens it forward to the image plane. Failed arcs are defined as those that do not produce additional images other than the original location, i.e., fail to produce the lensing multiplicity (see \autoref{sec:ippa}). The height of the bar represents the numbers of failed arcs by the model and the colors indicate the contribution from the training and new sets.}
\label{fig:fail_arcs}
\end{figure}

As can be seen in \autoref{fig:fail_arcs}, we find that the lens models are able to reproduce the majority of the $108$ multiple images, with a small fraction of systems that are not fully reproduced. Most of the models have less than $5$ failed arcs out of the $108$ arcs, e.g., overall there is $>95\%$ success rate and the models perform well in reproducing the lensing configuration. 

The JS-V3 model is able to produce a projection for each one of the $108$ arcs, while the W-V3 is the model with most failed arcs at $10$, most of them in the new set, in the north-east region of the cluster -- systems that also have high RMS. The failed arcs are shown in Figures~\ref{fig:loc_sps},~\ref{fig:loc_sps2},~\ref{fig:loc_sps3} as a square or circle with a line across the center, e.g., $\oslash$.

\section{Version 4 Lensing Models}
\label{sec:v4}

In 2017, the lensing teams repeated the vetting process for all the multiple images in \cname, including information on their spectroscopic redshifts now available from \citet{Caminha:17}. Using this new vetted list they computed the next version of lens models (V4).

Five of the teams: CATS (Cats-V4 and Cats-V4.1), Glafic (G-V4), Diego (D-V4 and D-V4.1), Sharon/Johnson (JS-V4), and Williams (W-V4), who had computed version 3 models, have new publicly available V4 models. In addition, some new groups: Caminha (Cam-V4) and Keeton (K-V4), have computed lens models and made then public through the \hff\ website. At the time of submission of this manuscript, there are no V4 models from the teams of Brada\v{c} and Zitrin. 

We have described the algorithms used by the CATS, Glafic, Diego, Sharon/Johnson, and Williams teams in \S\ref{subsec:lensing_models}. Caminha \citep{Caminha:17} used \texttt{LENSTOOL} \citep{Jullo:07,Jullo:09} which is the same algorithm used by CATS and Sharon/Johnson. Keeton uses GRAVLENS \citep{Keeton:11} which is also a parametric lens modeling algorithm.

\begin{figure*}
\center
\includegraphics[width=1\textwidth]{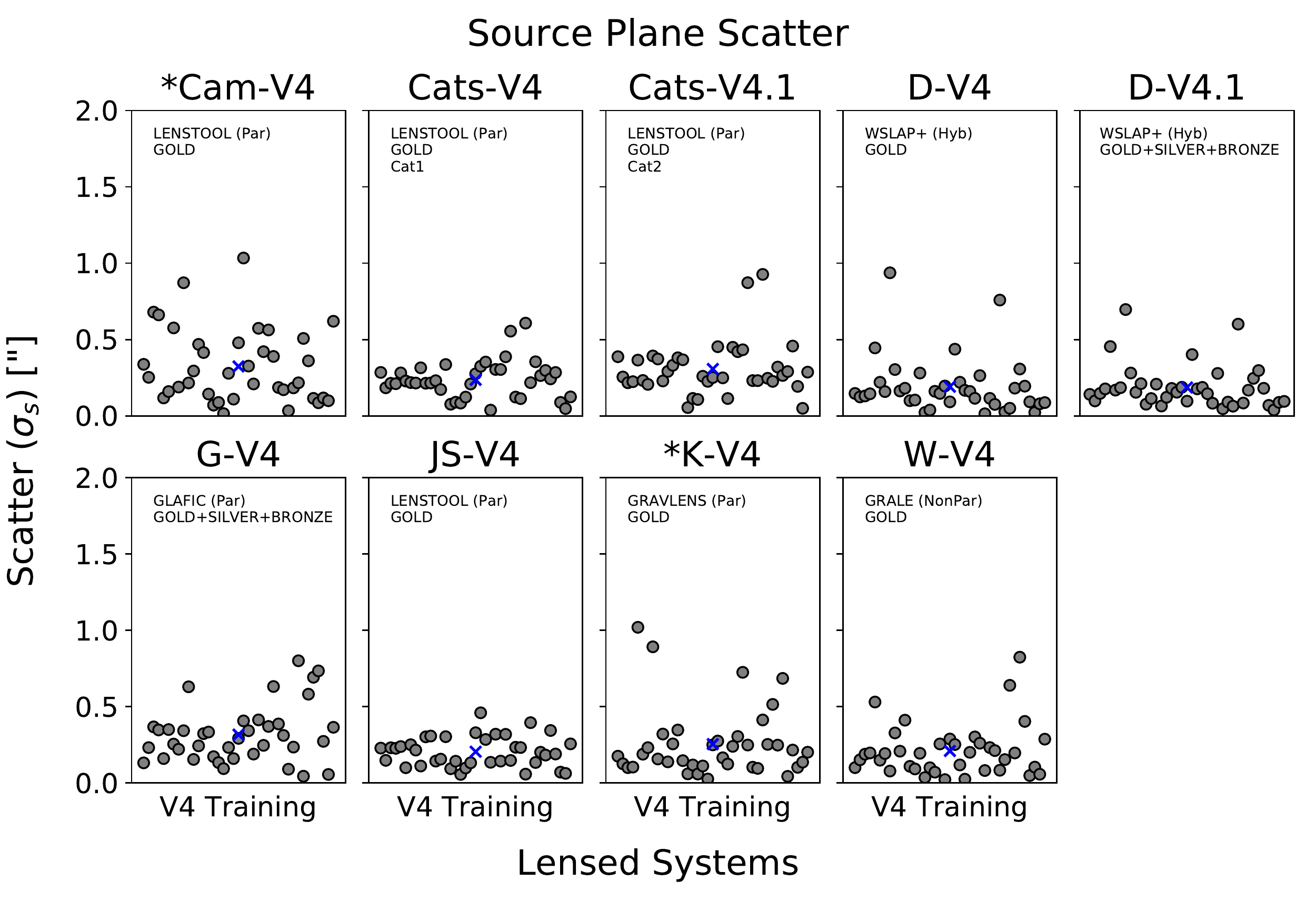}
\caption{Source plane RMS scatter plotted for each of the lensed systems using the version 4 publicly available lens models. All of the constraints are considered as training, since they were used as inputs when computing the new lens models. The blue ``x'' represents the mean of the distribution. In the text of each panel you can find the name of the algorithm used; in parenthesis if the algorithm is parametric (Par), non-parametric (NonPar), or hybrid (Hyb); the new constraints used, and additional comments to describe the difference between models from the same lensing group (two sets of galaxy cluster member catalogs, Cat1: \citealt{Grillo:15} and Cat2: \textcolor{blue}{Richard et al., in prep.}). The '*' in the name is to highlight the new teams that have made V4 lens models that did not compute a V3 lens models for \cname.}
\label{fig:v4}
\end{figure*}

We use the exact same constraints as in our analysis of the version 3 models (see \autoref{tab:data}) and the same methodology (see \S\ref{subsec:multiply-imaged}). We compute the source plane system RMS scatter ($\sigma_s$, \autoref{eq:system_RMS}) and plot it in \autoref{fig:v4} for each lensed system. In this case all of the constraints are considered training set, since they were used as constraints when computing the V4 lens models. In the CATS models, the different versions (V4 and V4.1) are not related to the quantity and quality of images, but indicate two different cluster-member galaxy catalogs (Cat1: \citealt{Grillo:15} and Cat2: \textcolor{blue}{Richard et al., in prep.}). In the models by Diego (\textcolor{blue}{Vega et al., in prep.}) two different versions 4 are related to the quality and quantity of constraints used to compute the models. 

In \autoref{fig:v4}, we observe that all of the lens models that are currently available on the Mikulski Archive for Space Telescopes (MAST) have a low RMS scatter and the distribution is similar for all of these models. Similarly, all the models have up to four failed arcs, with most having only $0-2$ failed arcs. That the RMS scatter is similar and low for all of the lens models and there are few failed arcs, demonstrates how well these lens models are doing in reproducing the lensing configuration.


\section{Summary and Conclusions}
\label{sec:conclusion}

In this work we evaluated the version 3, publicly available, \hff\ lensing models for \cname, by using the source plane RMS scatter and the image plane projection analysis as the metric. We have developed a method independent of all lensing algorithms used by the lensing teams to compute the source plane RMS scatter of the multiple images and predictions in the image plane. We compared the RMS scatter of the lens models using the training set - multiple images with spectroscopic redshifts that were known to the modelers at the time the models were computed - and the new arcs - multiple images that did not have spectroscopic redshifts at the time the models were computed. The new redshifts were published later by \citet{Caminha:17}. Here we summarize our findings:

\begin{itemize}
	\item When comparing the source plane RMS scatter of different models, we found that the new arcs have a higher RMS scatter compared to the training set in general for all the lens models (\autoref{fig:models_sps}). This is not surprising, since the models are tuned to reproduce the training systems.
    \item We find that C-V3 and C-V3.1 have a similar source plane scatter. The only difference between these two models is the number and quality of constraints used to compute the model: V3 used ``GOLD'' and ``SILVER'' constraints, while V3.1 used ``GOLD'', ``SILVER'', and ``BRONZE''. This indicates that parametric models do not necessarily benefit from adding more constraints with no spectroscopic redshift measurements when there are already hundreds of constraints that do have spectroscopic data, like in the case of the \hff\ clusters.
    \item We observe clear differences between the models W-V3 and W-V3.1. In the case of W-V3, where only ``GOLD'' and ``SILVER'' constraints were used, the majority of the RMS system scatter is fairly low with the exception of some outliers. On the other hand, W-V3.1 used all the ``GOLD'', ``SILVER'', and ``BRONZE'' arcs. Overall the source plane system RMS scatter is higher than that of W-V3. The observed differences between these two models tells us that this algorithm performs better when a higher fraction of the constraints used have high confidence and measured spectroscopic redshifts. 
    \item Comparing the RMS scatter against the source redshift (\autoref{fig:redshift_sps}), we found no trend with spectroscopic redshift. The only difference comes from the limited range of redshift from the training set (\autoref{fig:system_sps}) and the lower source plane RMS scatter of the training set compared to the new set.
    \item We computed the source plane scatter of system 26 (\autoref{fig:sps_arc26}) to assess the effect of the drastic change in the spectroscopic redshift of the system \citep{Caminha:17} from $z_{26} = 2.1851$ to $z_{26} = 3.2355$. We find that in all the models but one, W-V4.1, the RMS for this system decreases when calculated with the correct redshift.  In \hff, because there are so many constraints, one catastrophic failure is not going to break the model (just because of high scatter for this one system). In cases with a low number of constraints, however, it is significant \citep{Johnson:16}.
    \item We explore the spatial distribution of the system source plane RMS scatter against the location of the multiple-images in the image plane with \autoref{fig:loc_sps}, \ref{fig:loc_sps2} and \ref{fig:loc_sps3}. The models C-V3, C-V3.1, D-V3, G-V3, and JS-V3 have a uniform performance throughout the image plane. Z-LTM and Z-NFW, have relatively high RMS scatter in the south-west region of the image plane. Last, we observe drastic difference between the W-V3 and W-V3.1 models. Most of the W-V3 multiply-imaged systems have a low RMS scatter and are located in the middle and south-west regions of the field of view and three high-scatter systems are found in the north-east region of the image plane. On the other hand, in W-V3.1 the RMS scatter is uniformly high across the entire field of view. 
    \item Most of the parametric (C-V3, C-V3.1, G-V3, JS-V3, and Z-NFW), and hybrid (D-V3) models have a combined  average RMS scatter below $0\farcs5$. While the non-parametric models (B-V3 and W-V3.1) have a higher average RMS scatter. The W-V3 non-parametric model has a low average RMS scatter. Last the Z-LTM has the highest average RMS scatter overall.
    \item We have explored the predictive power of the lens models by comparing the distribution of the  new and training set system RMS scatter (\autoref{fig:delta_models_sps}). We find that the models which have the best predictive power are: C-V3, C-V3.1, D-V3, G-V3, JS-V3, and Z-NFW, which are found in the lower left part of \autoref{fig:delta_models_sps} and closest to the $1:1$ line.
    \item In the image plane, we find that the models correctly predict projections for most of the arcs, but there are some systems that fail to produce all the lensed images - we named these ``failed arcs''. We find that for most of the models, less than $5$ out of the 108 arcs failed (\autoref{fig:fail_arcs}). The model JS-V3 predicts projections for all of the multiple images while W-V3 has the most failed arcs at $10$, most of them found in the north-east region, near system 26, where the scatter is generally higher. This means that the lens models, regardless of the algorithm used, are very successful in reproducing most of the multiply-imaged systems.
    \item The new version 4 lens models for \cname\ have been made public by most of the lensing teams, as well as two new ones. We plot the source plane scatter in \autoref{fig:v4} for the models available at the time of the submission of this paper. We observe a similar low source plane RMS scatter distribution between all the lens models, demonstrating the progress in the quality of data, better characterization of \cname, and understanding of strong lens modeling systematics.
\end{itemize}

Recent studies that make extensive use of the HFF lens models, note differences between model outputs from the different algorithms (Livermore et al. 2017) particularly at high magnifications (Bouwens et al. 2017) highlighting a need for a direct comparison between lensing algorithms. Mahler et al. (2018) compared the effect of the addition of new spectroscopic redshifts on the uncertainties in the lensing mass reconstruction, magnifications, and redshift predictions. Our findings are consistent with their rms analysis concluding that lensing models benefit from having more spectroscopic measurements to the constraints.

This work complements other comparison projects like the one done by \citet{Priewe:17}, where the magnification of the lensing models was used as metric, and work done by \citet{Meneghetti:17}, where simulated clusters where used to evaluate the algorithms. While the RMS scatter of the source plane is not the only measure of performance of the models, it provides important insight. The analysis presented here forms a quantitative assessment of different models of real lenses, by confronting them with new observational lensing evidence. It  provides the community with means to understand where the strengths and differences of lens modeling algorithms lie; and with confidence that models of strong lensing clusters are becoming extremely powerful and reliable, allowing for their use as well-understood cosmic telescopes.  


\section{Acknowledgements}
\label{sec:acknowledgements} 

This material is based upon work supported by the National Science Foundation Graduate Research Fellowship Program. Any opinions, findings, and conclusions or recommendations expressed in this material are those of the author(s) and do not necessarily reflect the views of the National Science Foundation.

Some of the data presented in this paper were obtained from the Mikulski Archive for Space Telescopes (MAST). STScI is operated by the Association of the Universities for Research in Astronomy, Inc., under NASA contract NAS 5-26555. Support for MAST for non-\hst\ data are provided by the NASA Office of Space Science via grant NNX09AF08G and by other grants and contracts.

The authors would like to thank the anonymous referee for insightful suggestions that improved this manuscript. We thank all of the lensing teams and to all who contributed to the data that were used in this paper. In particular thanks to J. M. Diego and J. Vega for providing the correct lensing model and the coordinates of the constraints, and L. Williams and D. Coe for providing files that were not yet readily available on MAST. We thank M. Brada\v{c} for useful discussion. and  
 
\bibliographystyle{yahapj}
\bibliography{bibfile}

\appendix

\capstartfalse
\LongTables
\begin{deluxetable*}{cccccc}
\tablecolumns{6}
\tablecaption{List of lens model constraints that are used in this analysis.}
\tablehead{\colhead{ID} & \colhead{RA (deg)} & \colhead{DEC (deg)} & \colhead{$z$} 			  & \colhead{$z_{new}$} & \colhead{Training or New}}
\startdata
1.1 & 64.04075 & -24.061592 & 1.896 & 1.896 & Training \\
1.2 & 64.043479 & -24.063542 & 1.896 & 1.896 & Training \\
1.3 & 64.047354 & -24.068669 & 1.896 & 1.896 & Training \\
2.1 & 64.041165 & -24.061853 & 1.8925 & 1.895 & Training \\
2.2 & 64.043022 & -24.063024 & 1.8925 & 1.895 & Training \\
2.3 & 64.047487 & -24.068856 & 1.8925 & 1.895 & Training \\
3.1 & 64.03077 & -24.067128 & 1.9885 & 1.9894 & Training \\
3.2 & 64.035252 & -24.07099 & 1.9885 & 1.9894 & Training \\
3.3 & 64.041802 & -24.075731 & 1.9885 & 1.9894 & Training \\
4.1 & 64.03082 & -24.067231 & 1.9886 & 1.9887 & Training \\
4.2 & 64.035142 & -24.070968 & 1.9887 & 1.9887 & Training \\
4.3 & 64.041891 & -24.075856 & 1.9888 & 1.9887 & Training \\
5.1 & 64.032406 & -24.068414 & 2.0918 & 2.0948 & Training \\
5.2 & 64.032651 & -24.068666 & 2.0922 & 2.0948 & Training \\
5.3 & 64.03352 & -24.069451 & 2.0952 & 2.0948 & Training \\
5.4 & 64.0435551 & -24.07695762 & 2.088 & 2.0948 & Training \\
7.1 & 64.039798 & -24.063088 & 2.0854 & 2.0881 & Training \\
7.2 & 64.040667 & -24.063592 & 2.0854 & 2.0881 & Training \\
7.3 & 64.047102 & -24.071106 & 2.0854 & 2.0881 & Training \\
10.1 & 64.026053 & -24.077256 & 2.2982 & 2.2982 & Training \\
10.2 & 64.0284 & -24.079743 & 2.2982 & 2.2982 & Training \\
10.3 & 64.03669519 & -24.08390609 & 2.2982 & 2.2982 & Training \\
11.1 & 64.03925 & -24.070386 & 1.0054 & 1.0054 & Training \\
11.2 & 64.038296 & -24.06973 & 1.0054 & 1.0054 & Training \\
11.3 & 64.034233 & -24.066021 & 1.0054 & 1.0054 & Training \\
13.1 & 64.027567 & -24.072672 & 3.2226 & 3.2175 & Training \\
13.2 & 64.032161 & -24.075103 & 3.2226 & 3.2175 & Training \\
13.3 & 64.040353 & -24.081486 & 3.2226 & 3.2175 & Training \\
14.1 & 64.026233 & -24.074339 & 1.6324 & 1.6333 & Training \\
14.2 & 64.031042 & -24.078961 & 1.6344 & 1.6333 & Training \\
14.3 & 64.03583 & -24.081322 & 1.6344 & 1.6333 & Training \\
15.1 & 64.02686 & -24.075745 & 2.3355 & 2.3355 & Training \\
15.2 & 64.02944699 & -24.07858856 & 2.3355 & 2.3355 & Training \\
15.3 & 64.038234 & -24.082985 & 2.3355 & 2.3355 & Training \\
16.1 & 64.02406843 & -24.08089322 & 1.9644 & 1.9644 & Training \\
16.2 & 64.028334 & -24.084547 & 1.9644 & 1.9644 & Training \\
16.3 & 64.03161189 & -24.085762 & 1.9644 & 1.9644 & Training \\
17.1 & 64.029818 & -24.086364 & 2.2181 & 2.2182 & Training \\
17.2 & 64.028608 & -24.085981 & 2.2181 & 2.2182 & Training \\
17.3 & 64.02334 & -24.081581 & 2.2181 & 2.2182 & Training \\
23.1 & 64.044566 & -24.072098 & 2.0943 & 2.0932 & Training \\
23.2 & 64.039578 & -24.066631 & 2.0943 & 2.0932 & Training \\
23.3 & 64.034342 & -24.063742 & 2.091 & 2.0932 & Training \\
26.1 & 64.046452 & -24.0604 & 2.1851 & 3.2355 & Training \\
26.2 & 64.046963 & -24.060796 & 2.1851 & 3.2355 & Training \\
26.3 & 64.049083 & -24.062863 & 2.1851 & 3.2355 & Training \\
27.1 & 64.04813491 & -24.06695355 & 2.1067 & 2.1067 & Training \\
27.2 & 64.047469 & -24.066045 & 2.1067 & 2.1067 & Training \\
27.3 & 64.04221281 & -24.06054175 & 2.1067 & 2.1067 & Training \\
28.1 & 64.036506 & -24.067024 & 0.9377 & 0.9397 & Training \\
28.2 & 64.036838 & -24.067457 & 0.9377 & 0.9397 & Training \\
33.1 & 64.02840815 & -24.08299301 & \nodata & 5.365 & New \\
33.2 & 64.03505002 & -24.08549952 & \nodata & 5.365 & New \\
33.3 & 64.02298453 & -24.07726749 & \nodata & 5.365 & New \\
34.1 & 64.029254 & -24.073289 & \nodata & 5.106 & New \\
34.2 & 64.030798 & -24.07418 & \nodata & 5.106 & New \\
35.1 & 64.037492 & -24.083636 & \nodata & 3.4909 & New \\
35.2 & 64.029418 & -24.079861 & \nodata & 3.4909 & New \\
35.3 & 64.024937 & -24.075016 & \nodata & 3.4909 & New \\
38.1 & 64.033625 & -24.083178 & \nodata & 3.4406 & New \\
38.2 & 64.031255 & -24.081905 & \nodata & 3.4406 & New \\
38.3 & 64.022701 & -24.074589 & \nodata & 3.4406 & New \\
44.1 & 64.045259 & -24.062757 & \nodata & 3.2885 & New \\
44.2 & 64.041543 & -24.059997 & \nodata & 3.2885 & New \\
44.3 & 64.049237 & -24.068168 & \nodata & 3.2885 & New \\
47.1 & 64.026328 & -24.076694 & \nodata & 3.2526 & New \\
47.2 & 64.028329 & -24.078999 & \nodata & 3.2526 & New \\
48.1 & 64.035489 & -24.084668 & \nodata & 4.1218 & New \\
48.2 & 64.029244 & -24.081802 & \nodata & 4.1218 & New \\
48.3 & 64.023416 & -24.076122 & \nodata & 4.1218 & New \\
49.1 & 64.033944 & -24.074569 & \nodata & 3.871 & New \\
49.2 & 64.040175 & -24.079864 & \nodata & 3.871 & New \\
51.1 & 64.04013543 & -24.08029748 & \nodata & 4.1032 & New \\
51.2 & 64.03366471 & -24.0747629 & \nodata & 4.1032 & New \\
51.3 & 64.02663527 & -24.07047437 & \nodata & 4.1032 & New \\
55.1 & 64.035233 & -24.064726 & \nodata & 3.2922 & New \\
55.3 & 64.038514 & -24.065965 & \nodata & 3.2922 & New \\
58.1 & 64.025187 & -24.073582 & \nodata & 3.0773 & New \\
58.2 & 64.03773 & -24.08239 & \nodata & 3.0773 & New \\
58.3 & 64.030481 & -24.07922 & \nodata & 3.0773 & New \\
67.1 & 64.038075 & -24.082404 & \nodata & 3.1103 & New \\
67.2 & 64.025451 & -24.073651 & \nodata & 3.1103 & New \\
67.3 & 64.030363 & -24.079019 & \nodata & 3.1103 & New \\
22\_D15.1 & 64.03448151 & -24.06695655 & \nodata & 3.2215 & New \\
22\_D15.2 & 64.034181 & -24.066489 & \nodata & 3.2215 & New \\
22\_D15.3 & 64.034006 & -24.066447 & \nodata & 3.2215 & New \\
32\_D15.1 & 64.045119 & -24.072336 & \nodata & 3.2882 & New \\
32\_D15.2 & 64.040081 & -24.06673 & \nodata & 3.2882 & New \\
2\_C16.a & 64.050865 & -24.066538 & \nodata & 6.1452 & New \\
2\_C16.b & 64.048179 & -24.062406 & \nodata & 6.1452 & New \\
2\_C16.c & 64.043572 & -24.059004 & \nodata & 6.1452 & New \\
6\_C16.a & 64.047781 & -24.070169 & \nodata & 3.6065 & New \\
6\_C16.b & 64.043657 & -24.064401 & \nodata & 3.6065 & New \\
6\_C16.c & 64.037679 & -24.060756 & \nodata & 3.6065 & New \\
17\_C16.a & 64.040489 & -24.07838 & \nodata & 3.9663 & New \\
17\_C16.b & 64.035107 & -24.073864 & \nodata & 3.9663 & New \\
17\_C16.c & 64.027171 & -24.068224 & \nodata & 3.9663 & New \\
22\_C16.b & 64.030997 & -24.077173 & \nodata & 3.923 & New \\
22\_C16.c & 64.027127 & -24.073572 & \nodata & 3.923 & New \\
23\_C16.a & 64.035668 & -24.07992 & \nodata & 2.5425 & New \\
23\_C16.b & 64.032638 & -24.078508 & \nodata & 2.5425 & New \\
33\_C16.a & 64.032017 & -24.08423 & \nodata & 5.9729 & New \\
33\_C16.b & 64.030821 & -24.083697 & \nodata & 5.9729 & New \\
34\_C16.b & 64.027632 & -24.082609 & \nodata & 3.9228 & New \\
34\_C16.c & 64.023731 & -24.078477 & \nodata & 3.9228 & New \\
35\_C16.a & 64.033681 & -24.085855 & \nodata & 5.639 & New \\
35\_C16.b & 64.028654 & -24.08424 & \nodata & 5.639 & New \\
35\_C16.c & 64.022187 & -24.077559 & \nodata & 5.639 & New

\enddata
\tablenotetext{}{Each image is identified by its ID, Right Ascension (RA), and Declination (DEC) in degrees (J2000). The next columns are the spectroscopic redshifts ($z$) available in 2015 \citep{Jauzac:14,Richard:14,Treu:15}, and the newly available spectroscopic redshifts ($z_{new}$) from \citet{Caminha:17}. The last column indicates whether the system is considered part of the training set or new set.}
\label{tab:data}
\end{deluxetable*}
\capstarttrue

\end{document}